\documentclass[prl,aps,twocolumn,showpacs,superscriptaddress,footnoteinbib]{revtex4-1}

\usepackage{graphicx}
\usepackage{dcolumn}
\usepackage{bm}
\usepackage{amssymb,amsmath}
\usepackage{sidecap}
\usepackage{wrapfig}
\usepackage{natbib}

\usepackage{graphicx}
\usepackage{leftidx}
\usepackage{color}

\usepackage{bbold} 

\usepackage{wasysym} 

\usepackage{stackrel}

\usepackage{soul,xcolor}
\setstcolor{red}

\newcommand{\be}{\begin{equation}}
\newcommand{\ee}{\end{equation}}
\newcommand{\ba}{\begin{align}}
\newcommand{\ea}{\end{align}}
\newcommand{\sysb}{\left\{\begin{array}}
\newcommand{\syse}{\end{array}\right.}
\newcommand{\baa}{\begin{array}}
\newcommand{\eaa}{\end{array}}
\newcommand{\bs}{\begin{split}}
\newcommand{\es}{\end{split}}

\newcommand{\matb}{\left(\begin{array}}
\newcommand{\mate}{\end{array}\right)}

\newcommand{\mal}{\mathcal}
\newcommand{\rmd}{{\rm{d}}}

\newcommand{\rme}[1]{{\rm{e}}^{#1}}
\newcommand{\mand}{\quad\text{ and }\quad}

\newcommand{\wt}{\widetilde}

\newcommand{\id}{\mathbb{1}}

\newcommand{\ha}{\frac{1}{2}}

\newcommand{\sdim}[1]{\lqq #1 \rqq}

\newcommand{\lt}{\left(}
\newcommand{\rt}{\right)}
\newcommand{\lqq}{\left[}
\newcommand{\rqq}{\right]}
\newcommand{\lan}{\left\langle}
\newcommand{\ran}{\right\rangle}

\newcommand{\av}[1]{\lan #1 \ran}

\newcommand{\set}[1]{\left\{  #1  \right\}}

\newcommand{\ket}[1]{\left| #1 \ran}
\newcommand{\bra}[1]{\lan #1 \right|}

\newcommand{\comm}[2]{\left[ #1, #2 \right]}
\newcommand{\acomm}[2]{\left\{ #1, #2 \right\}}

\newcommand{\changer}[1]{\textcolor{black}{#1}}

\newcommand{\comma}{\quad , \quad}

\usepackage{amsthm}

\begin{document}

\title{Epidemic dynamics in open quantum spin systems}

\author{Carlos P\'erez-Espigares}
\affiliation{School of Physics and Astronomy, University of Nottingham, Nottingham, NG7 2RD, United Kingdom
and Centre for the Mathematics and Theoretical Physics of Quantum Non-equilibrium Systems,
University of Nottingham, Nottingham NG7 2RD, UK}
\author{Matteo Marcuzzi}
\affiliation{School of Physics and Astronomy, University of Nottingham, Nottingham, NG7 2RD, United Kingdom
and Centre for the Mathematics and Theoretical Physics of Quantum Non-equilibrium Systems,
University of Nottingham, Nottingham NG7 2RD, UK}
\author{Ricardo Guti\'errez}
\affiliation{School of Physics and Astronomy, University of Nottingham, Nottingham, NG7 2RD, United Kingdom
and Centre for the Mathematics and Theoretical Physics of Quantum Non-equilibrium Systems,
University of Nottingham, Nottingham NG7 2RD, UK}
\affiliation{Complex Systems Group, Universidad Rey Juan Carlos, 28933 M\'ostoles, Madrid, Spain.}
\author{Igor Lesanovsky}
\affiliation{School of Physics and Astronomy, University of Nottingham, Nottingham, NG7 2RD, United Kingdom
and Centre for the Mathematics and Theoretical Physics of Quantum Non-equilibrium Systems,
University of Nottingham, Nottingham NG7 2RD, UK}

\begin{abstract}
We explore the non-equilibrium evolution and stationary states of an open many-body system which displays epidemic spreading dynamics in a classical and a quantum regime. Our study is motivated by recent experiments conducted in strongly interacting gases of highly excited Rydberg atoms where the facilitated excitation of Rydberg states competes with radiative decay. These systems approximately implement open quantum versions of models for population dynamics or disease spreading where species can be in a healthy, infected or immune state. We show that in a two-dimensional lattice, depending on the dominance of either classical or quantum effects, the system may display a different kind of non-equilibrium phase transition. We moreover  discuss the observability of our findings in laser driven Rydberg gases with particular focus on the role of long-range interactions.
\end{abstract}

\maketitle

\textit{Introduction ---} Cold atoms and ions are versatile platforms for the exploration of non-equilibrium physics. Recent examples include studies on creation and dynamics of quasi-particles \cite{schirotzek2009,jurcevic2014}, spreading of entanglement and correlations \cite{islam2015,fukuhara2015,kaufman2016}, as well as many-body localization in disordered systems \cite{choi2016,schreiber2015,smith2016}. In particular, so-called Rydberg gases with their strong long-range interactions \cite{Ryd-QI,Rydberg2} permit the exploration of open and closed many-body physics \cite{schauss2012,Schempp_2014,Urvoy_2015,Weber_2015,Helmrich2016}, with recent experiments probing non-equilibrium dynamics \cite{Helmrich2016,Valado_2016,letscher2017bistability,goldschmidt2016anomalous}, phase transitions \cite{carr2013,malossi2014,gutierrez2016,ding2016} and disorder-induced localization phenomena \cite{marcuzzi2017}.

An intriguing aspect is that in Rydberg gases one can achieve control over the relative strength of quantum fluctuations and classical noise \cite{Rydberg2}. This permits the exploration of dynamical phenomena in settings which can be regarded as quantum generalizations of classical non-equilibrium systems \cite{hinrichsen2000,odor2004}. A recent example is a quantum version of the so-called contact process \cite{Marcuzzi_2016,buchhold2017}, a simple stochastic model for population dynamics featuring a non-equilibrium phase transition \cite{henkel2008non} whose character changes drastically when moving from a purely-classical to a quantum regime. 

\begin{figure}[t!]
  \includegraphics[width=\columnwidth]{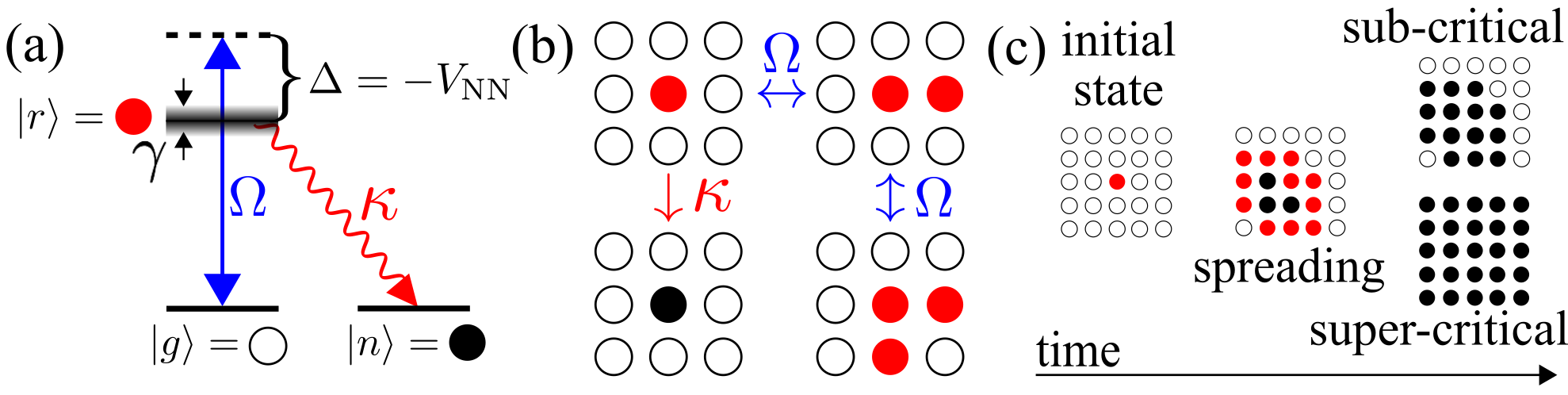}
  \caption{(a) Atoms on a square lattice are coherently excited from the ground state $|g \rangle$ to a Rydberg state $|r \rangle$ with a laser with Rabi frequency $\Omega$. External noise broadens the state $|r \rangle$ (width $\gamma$) which decays to a third state $|n \rangle$ at rate $\kappa$. The laser is off-resonant with a detuning $\Delta$ that compensates the nearest-neighbor interaction $V_\mathrm{NN}$ ($V_\mathrm{NN}-\Delta=0$). (b) The dominant processes are facilitation (top row and right column) and decay (left column). (c) An initial seed leads to the formation of clusters of Rydberg states (infected sites) which can either be converted to ground state atoms (healthy sites) or decay to the immune state $|n \rangle$. The relative strength of the dephasing rate $\gamma$ with respect to $\Omega$ determines the nature of the transition. At fixed $\gamma$, depending on the ratio $\Omega/\kappa$ the stationary state is either an ever-expanding infection leaving a macroscopic fraction of immune sites (super-critical) or an infection that dies and leaves a lattice partially (not macroscopically) filled (sub-critical). }
\label{fig:1}
\end{figure}

In this work we shed light on the collective dynamics of an open quantum system generalization of a \emph{general epidemic process} (GEP) \cite{Grassberger1983}, belonging to the dynamic  percolation universality class \cite{cardy1983field,cardy1985epidemic,Janssen1985,henkel2008non}. 
\changer{In a Rydberg system, a similar dynamics can be expected by considering atoms}
with three relevant states, which can be labeled as ``healthy'', ``infected'' and ``immune'', where infected sites have the ability to infect their healthy neighbors, or heal and become immune. A scenario similar to this has been recently realized and studied experimentally in \cite{Helmrich2016}, where a connection to the GEP was conjectured. \changer{The scope of this work is not to propose a quantum simulation protocol for GEP, but to demonstrate that}, under dominant classical noise, the system follows the same phenomenology as the GEP and undergoes a continuous transition between two phases: one where the contagion starting from an initially infected site is unable to percolate throughout the system, and one where the initial infected site triggers a self-sustaining wavefront (an ``outbreak'') which covers an extensive portion of the system and leaves behind a trail of immunized sites.
\changer{In the quantum regime, a mean-field treatment suggests that the density of immune sites displays a sequence of jumps resulting from the presence of multiple wavefronts.}

\textit{Model ---} We consider atoms with three internal states: a ground state $|g\rangle$ (healthy), a Rydberg state $|r \rangle$ (infected) and a second stable state $|n\rangle$ (immune). These states are coupled as depicted in Fig. \ref{fig:1}a: $|g\rangle$ is excited to $|r \rangle$ via a laser with Rabi frequency $\Omega$ and detuning $\Delta$ and the state $|r \rangle$ decays radiatively into $|n \rangle$ at rate $\kappa$. \changer{Note that this implicitly assumes that the decay from $|r \rangle$ does not proceed via long-lived intermediate states. Radiative decay from $|r \rangle$ to $|g \rangle$ is neglected for simplicity (see \cite{SM} for details).}
The atoms are placed on a two-dimensional square lattice with $L$ sites and spacing $a$, one per site (see e.g. experiments in Refs.~\cite{Barredo2015,Schauss_2015}).

Collective behavior emerges when the probability for an atom to undergo the transition $|g \rangle\rightarrow|r \rangle$ (infection of a healthy site) depends on the state of its neighbors. For Rydberg atoms this is achieved by enforcing the so-called ``facilitation'' (or ``anti-blockade'') condition \cite{Ates_2007,Amthor_2010,Gartner_2013,schonleber2014,Lesanovsky_2014,Urvoy_2015,Valado_2016}. Here, the detuning $\Delta$ of the excitation laser is set to compensate the interaction $V_\mathrm{NN}$ between neighboring atoms, which makes the transition $|g \rangle\rightarrow|r \rangle$ resonant provided a neighbor is already in state $|r \rangle$ (Fig.~\ref{fig:1}(a)). This situation has already been explored in a two-level setup \cite{Schempp_2014,Urvoy_2015,Valado_2016,thaicharoen2015}, and very recently also in the considered three-state setting \cite{Helmrich2016}. We remark that it is crucial for infection to only occur locally, i.e.~in the neighborhood of an already infected site. In our case, it is therefore important that the interactions decay sufficiently fast with the distance. A variety of different behaviors are known to emerge in systems where they do not \cite{bouchet2005,chatterjee2013,fischer2015,fasshauer2016}.

We now consider a minimal model of the resulting many-body dynamics in which the density matrix $\rho$ of the system evolves under a Lindblad master equation \cite{Breuer2002} $\partial_t\rho=-i\left[H,\rho\right]+ \sum^L_{k=1}\left[\kappa\mathcal{L}\left(\left|n\right>_k\!\left<r\right|\right)\rho + \gamma \mathcal{L}\left(r_k\right)\rho\right]$,
whose terms are sketched in Fig.~\ref{fig:1}(b).
The coherent evolution is governed by the Hamiltonian $H=\Omega\sum_k \Pi_k \sigma_k^x$ where $\sigma_k^x=\left|g\right>_k\!\left<r\right|+\left|r\right>_k\!\left<g\right|=\sigma^-_k+\sigma^+_k$, and $\Pi_k$ is a projector onto the subspace in which exactly one of the nearest neighbors of $k$ is in state $\left|r\right>$. Denoting the set of nearest neighbors of $k$ by $\Lambda_k$, it reads
\begin{eqnarray}
  \Pi_k\!=\!\sum_{l\in\Lambda_k}\!r_{l}\!\prod_{m \in \Lambda_k\backslash\set{l}}\!(1\! -\! r_m) =\!\sum_{l\in\Lambda_k}\!r_{l} + \ldots,  \label{eq:pik}
\end{eqnarray}
with $r_k=\left|r\right>_k\!\left<r\right|$,  $g_k=\left|g\right>_k\!\left<g\right|$ and $n_k=\left|n\right>_k\!\left<n\right|$. The dots denote higher-order terms in the operators $r_k$. This projector 
constrains infection to occur only in the neighborhood of a single infected site. The dissipative dynamics is described by Lindblad terms $\mathcal{L}\left(J\right)\rho=J\rho J^\dagger - \{J^\dagger J,\rho\}/2$. The first describes decay from $|r \rangle$ to $|n \rangle$ at rate $\kappa$, the second dephasing of quantum coherences at rate $\gamma$. Controlling the dephasing strength, which is achievable by modifying the excitation laser linewidth or the temperature of the atoms, allows switching between classical and quantum regimes \cite{Rydberg2,levi2016}. In the following, the initial state is always a single atom in state $\ket{r}$ (infected) in the center of the lattice and all the others in state $\ket{g}$ (healthy).

\textit{Classical regime ---} We first consider the regime of strong dephasing $\gamma \gg \Omega$. Here an effective dynamics can be defined for the diagonal of the density matrix $\mu_{ij} = \delta_{ij} \rho_{ii}$ in the $\ket{r,g,n}$ basis and the corresponding (classical) master equation reads \cite{Degenfeld14, marcuzzi2014}
\begin{align}
  \partial_t\mu=\sum_k\! \left[\alpha \Pi_k\left(\mathcal{L}(\sigma^+_k)+\mathcal{L}(\sigma^-_k)\right)+\kappa\mathcal{L}\left(\left|n\right>_k\!\left<r\right|\right)\right]\mu,\label{eq:stochastic_dynamics}
\end{align}
with $\alpha=4\Omega^2/\gamma$. This means that atoms undergo incoherent state changes from $|g\rangle$ to $|r\rangle$ and vice versa, with a rate conditioned on their local neighborhood. Furthermore, decay from $|r\rangle$ to $|n\rangle$ is possible. The process \eqref{eq:stochastic_dynamics} is similar to a GEP but differs from it by (i) the presence of facilitated transitions $\ket{r} \to \ket{g}$ and (ii) for the stricter constraint that a \emph{single} neighboring infected site is required for facilitation, whereas in the GEP the infection rate is proportional to the number of infected neighbors. 

In a GEP, for fixed $\kappa$, if the facilitation rate lies below its critical value, the initial infection is unable to propagate and the density of immunes $N=\sum_k\langle n_k \rangle/L$ ($L$ being the number of sites) vanishes in the thermodynamic limit. Conversely, above the critical point there is a finite probability for the infection to percolate (see Fig.~\ref{fig:1}(c)), propagating as a single travelling wavefront and leaving behind a finite fraction of immune sites $N > 0$ \cite{Grassberger1983} (i.e.~a single outbreak takes place). 

\begin{figure}[t!]
  \includegraphics[width=\columnwidth]{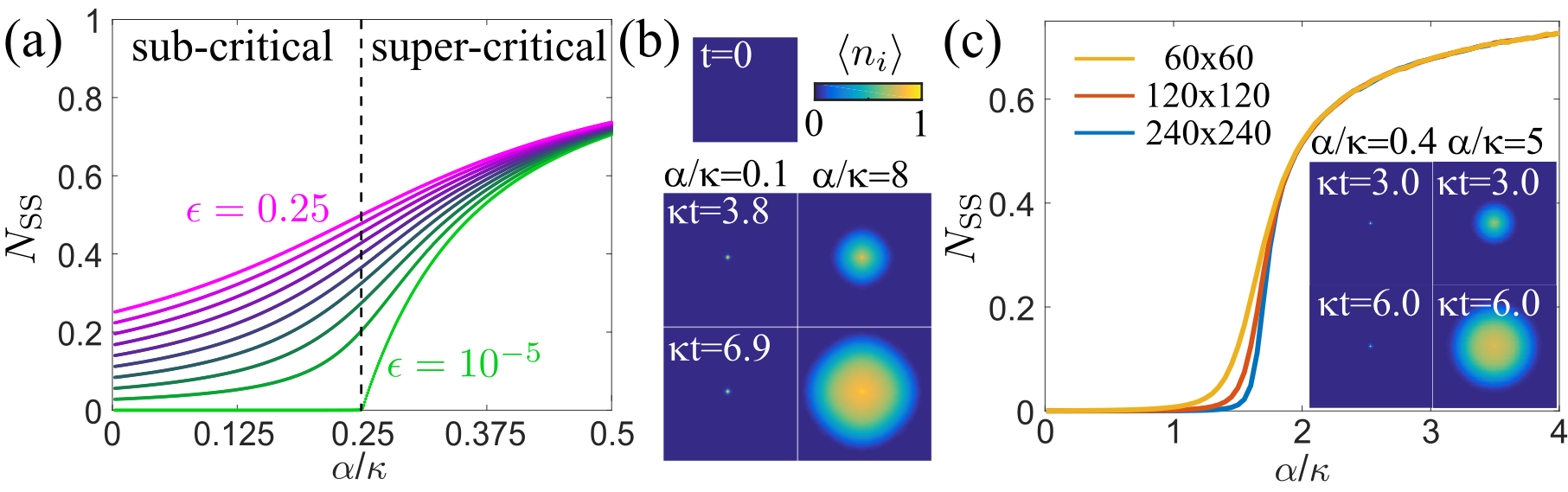}
  \caption{(a) Stationary state of the mean-field equations (\ref{eq:classical_mf}). Numerical solution after time $\kappa t=500$ for different initial conditions: $R(t=0)=\epsilon$, $G(t=0)=1-\epsilon$ and $N(t=0)=0$. (b) Evolution of $\langle n_i\rangle$ on a $L=51\times 51$ lattice starting at $t=0$ from a single site in state $|r\rangle$ located at the center of the lattice. In the super-critical regime the epidemics spreads through the lattice leaving a region of immune sites behind, while in the sub-critical regime the spreading soon halts. (c) The Monte Carlo simulation of (\ref{eq:stochastic_dynamics}) displays a continuous phase transition as well, with order parameter profiles becoming sharper the larger the system size. Snapshots are averaged over $10^4$ realizations and display the expected qualitative features.}
\label{fig:2}
\end{figure}

We start from a uniform mean-field approximation where we neglect the higher order terms in \eqref{eq:pik}, effectively relaxing (ii). Introducing the quantities $R=\sum_k\langle r_k \rangle/L$ and $G=\sum_k\langle g_k \rangle/L$, this yields
\begin{eqnarray}
 \partial_t G &=& -4\alpha R(G-R), \quad\partial_t N = \kappa R\nonumber\\
 \partial_t R &=& 4\alpha R(G-R)-\kappa R.\label{eq:classical_mf}
\end{eqnarray}
Analogously to what is found for the GEP \cite{Grassberger1983}, these equations feature a constant of motion $\partial_t(\log(R-G+\kappa/8\alpha)+8N\alpha/\kappa)=0$ which permits the determination of the stationary phase diagram for different initial conditions, shown in Fig.~\ref{fig:2}(a)  \cite{SM}. In a uniform approximation, the closest initial condition to the one we start from is a vanishingly small density of infections ($R(t=0) = \epsilon \to 0^+$) and in this limit the two phases (which are illustrated in Fig.~\ref{fig:1}(c)) can be clearly identified, separated by a critical point at $\alpha_c = \kappa /4$. The non-uniform mean-field dynamics is shown in Fig.~\ref{fig:2}(b) and highlights the absence/presence of an outbreak in the two phases.

We have then performed continuous-time Monte Carlo simulations of the classical master equation (\ref{eq:stochastic_dynamics}). In Fig.~\ref{fig:2}(c) we show the stationary density $N_\mathrm{SS} = \lim_{t \to \infty} N(t)$ for different system sizes and observe a sharpening crossover from a vanishing to a finite-valued phase when increasing $L$. The dynamics shows the expected GEP behavior: for $\alpha < \alpha_c \approx 1.71\, \kappa$ the process fails to percolate and no outbreak is produced. In the supercritical phase, instead, there is a finite probability of a single outbreak immunizing a macroscopic portion of the system. A finite-size scaling analysis performed with the tabulated critical exponents for the GEP provides an excellent collapse of the curves \cite{SM}, establishing the connection on firm grounds. The full constraint given by \eqref{eq:pik} leads to the same scaling behavior near the critical point. This could be expected since the differences between our process and the GEP are of higher order in the density of infected sites, which vanishes close to the critical point. More rigorously, one can show that these terms are irrelevant in a renormalization-group sense \cite{SM}. 

\textit{Rydberg gases ---}  We now discuss the observability of this physics in Rydberg gases interacting with a power-law potential $V(r)=C_\beta/r^\beta$ (here we consider the van der Waals case $\beta = 6$) under anti-blockade conditions. As shown in Refs.~\cite{ates2007,Lesanovsky_2014,marcuzzi2014,marcuzzi2015}, in this case the constraint $\Pi_k$ is replaced by the operator
$
\Gamma_k = 1/(1+R^{2\beta}(1-\sum_{m\ne k}\frac{r_m}{| {\bf {x}}_k-{\bf {x}}_m|^\beta})^2),
$
where $R=(2 C_\beta/a^\beta \gamma)^{1/\beta}$ is the so-called dissipative blockade radius and ${\bf {x}}_k$ denotes the position of the $k$-th atom. The lattice spacing $a$ is taken as the facilitation distance: if just one nearest neighbor of site $k$ is infected, the rate is maximized, i.e. $\Gamma_k = 1$, whereas other configurations lead to its suppression. In this context, two main effects alter the physics of the previous model: the possibility of unfacilitated infection ($\ket{g} \leftrightarrow \ket{r}$ in the absence of infected neighbors) and the fact that the interactions extend beyond nearest neighbors \cite{gutierrez2016}. To gain insight into their role, we exploit the rapid decay of the tails of $V(r)$ to truncate the interaction beyond a distance of two lattice sites, 
\begin{equation}
\Gamma_k^{-1} \approx 1+R^{2\beta}\left[1-\sum_{l \in \Lambda_k} r_{l} - \eta \sideset{}{'}\sum_{m}\frac{r_m}{| {\bf {x}}_k-{\bf { x}}_m|^\beta}\right]^2,
\label{rydberg_rate}
\end{equation}
where the primed sum runs only over $m$ such that $1<| {\bf { x}}_k-{\bf {x}}_m| \leq 2$. Here we have introduced a parameter $\eta$, which allows us to control the strength of the `long-range' part. While this parametrization is used here for convenience, in practice potential shaping techniques can be applied to modify and possibly suppress the potential tails, see e.g.~Refs.~\cite{buchler2007,sevinccli2014,marcuzzi2015}.

\begin{figure}[t!]
  \includegraphics[width=\columnwidth]{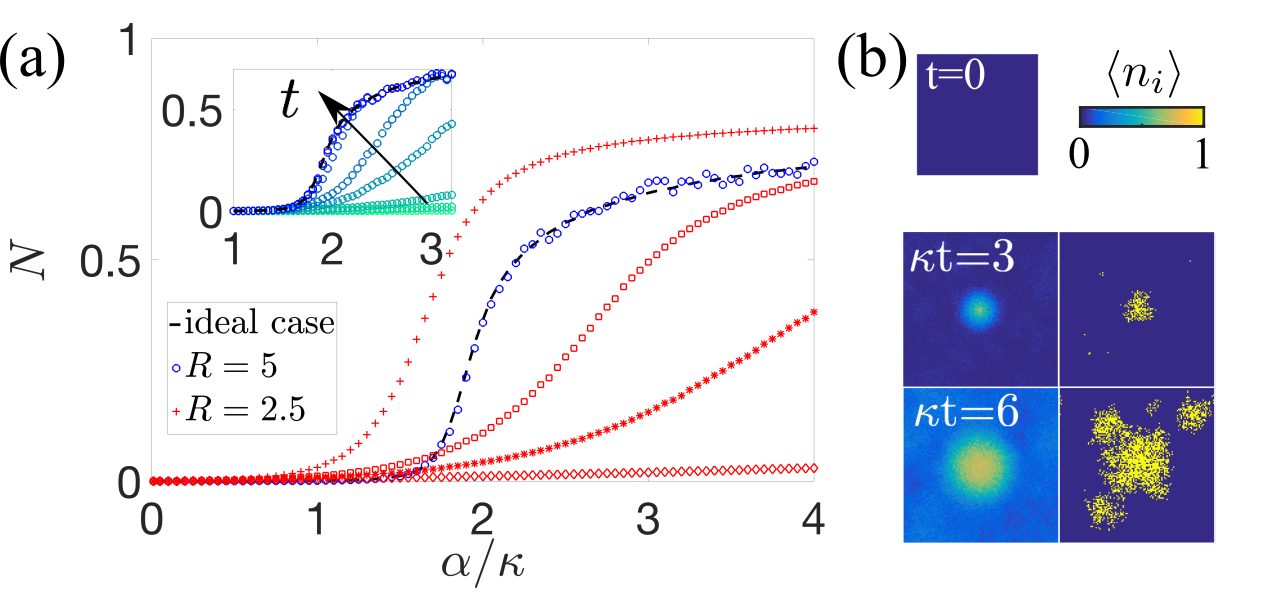}
  \caption{(a) Mean density of immune states in the ideal case with rates $\alpha \Pi_k$ \eqref{eq:pik} (black dashed line), and in the Rydberg case \eqref{rydberg_rate} at time $\kappa t=100$ for $R=5$ and $\eta = 0$ (blue circles) and for $R = 2.5$ and different values of $\eta=0,0.05,0.1,1$ (red symbols from top to bottom). The Rydberg model prediction matches the idealized model's when the effects of the potential tails and of spontaneous excitation can be neglected. Inset: Mean density of immune states using the Rydberg constraint \eqref{rydberg_rate} with $R=5$, $\beta=6$ and $\eta=0$ for increasing values of time $\kappa t=3, 6, 10, 20, 30, 60, 90$ and $100$. The ideal case with constraints $ \Pi_k$ (black dashed line) is again included for comparison. (b) Evolution of the density of immune states averaged over $100$ realizations (left column) and for a single realization (right column) on a $L=120\times 120$ lattice for $\alpha=5\kappa$, $R=2.5$ and $\eta=0$ at two different values of $\kappa t$.}
\label{fig:4}
\end{figure}

First, we consider $\eta = 0$ (nearest-neighbors interactions only), where $\Gamma_k$ is well approximated by the constraint $\Pi_k$ [Eq. \eqref{eq:pik}], provided that $R > 1$. Unfacilitated (spontaneous) infection can now occur at a rate $\alpha \Gamma_\mathrm{spont} = \alpha/(1 + R^{2\beta}) > 0$. Albeit rare for $R \gg 1$, these processes dramatically alter the stationary-state properties of the system, invariably leading to $N_\mathrm{SS} = 1$. In the renormalization-group language, the spontaneous processes constitute a relevant perturbation.  Nevertheless, for sufficiently large $R$ a timescale separation occurs: the outbreak follows the phenomenology observed for the idealized case up to times on the order of $\kappa t \sim (1+R^{2\beta})/ (L \alpha/\kappa) $, which is an underestimate of the mean waiting time of the Poisson process producing spontaneous infection. This is illustrated in Fig.~\ref{fig:4}(a), where we show the stationary density of immune states $N_\mathrm{SS}$ of the idealized process (rate function $\alpha \Pi_k$), together with the density $N(t)$ at time $\kappa t = 100$ for the Rydberg rates with $R=5$, $\beta=6$ and $\eta=0$. These curves display remarkable agreement showing that this phase transition in fact underlies the transient Rydberg dynamics. In the inset of Fig.~\ref{fig:4}(a) we moreover show how the characteristic sigmoidal profile remains stable for a long period of time. In Fig.~\ref{fig:4}(b) we show that outbreaks are visible for some time, but spontaneous infection leads to a rising ``background" which eventually overcomes the epidemic process and immunizes the entire system.

Long-range interactions ($\eta > 0$) counteract the formation of large clusters of infected and immune sites. For instance, an isolated atom in state $|r\rangle$ can infect, say, the site directly below it at the maximal rate, since the nearest-neighbor interaction compensates the laser detuning. To infect a third one at the right of the former, however, one has to include the additional shift due to next-nearest-neighbor interactions, which brings the atomic transition off resonance, hindering the propagation.
We illustrate this by showing in Fig.~\ref{fig:4}(a) the density of immune states at time $\kappa t=100$ for $R=2.5$ and different values of $\eta$ (all red symbols). As $\eta$ is increased, it becomes more difficult for the process to spread. This however is a lattice effect; in dense atomic clouds the transition may reappear. 
\changer{This is suggested by recent experimental works \cite{Urvoy_2015,Valado_2016,Helmrich2016,gutierrez2016} that reveal collective phenomena in the presence of thermal motion.}

\begin{figure}[t!]
  \includegraphics[width=\columnwidth]{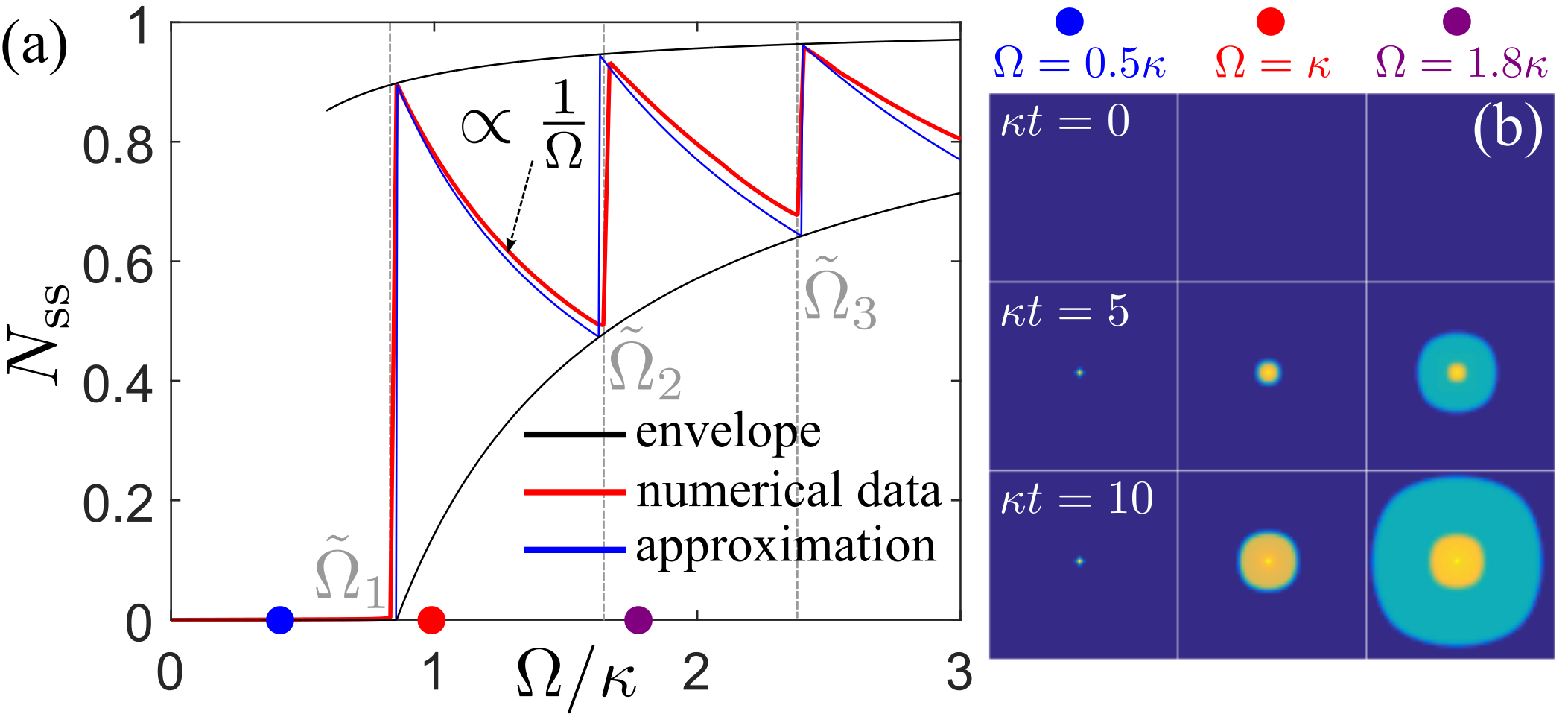}
  \caption{(a) Stationary state of the mean-field equations (\ref{eq:quantum_mf}). We show the numerical solution for the immune density at time $\kappa t=500$ starting at $t = 0$ from a single infected site located at the center of a $51 \times 51$ lattice. As the driving parameter $\Omega/ \kappa$ is increased, a recurrent structure of jumps is seen. (b) Snapshots of the numerical mean-field evolution of the same process for three values of $\Omega$ showing how every jump in $N_{\rm{SS}}$ is associated to the appearance of a new outbreak.}  
\label{fig:3}
\end{figure}

\textit{Quantum regime ---} We now set the dephasing rate $\gamma=0$ and study what we refer to as the ``quantum case''. We describe the dynamics via the mean-field equations
\begin{eqnarray}
  \partial_t G_k &=& - \Omega \overline{R}_k \, \Sigma_k, \  \partial_t R_k = -\Omega \overline{R}_k \, \Sigma_k - \kappa R_k ,\nonumber\\
    \partial_t N_k &=& \kappa R_k, \  \partial_t \Sigma_k = -2\Omega \overline{R}_k (R_k-G_k)-\frac{\kappa}{2}\Sigma_k. \label{eq:quantum_mf}
\end{eqnarray}
Here, $R_k \, (G_k, N_k) = \av{r_k \, (g_k, n_k)} $ and $\Sigma_k = \langle \sigma^y_k \rangle$ with $\sigma_k^y=i\left|g\right>_k\!\left<r\right|-i\left|r\right>_k\!\left<g\right|$. Additionally, $\overline{R}_k = \sum_{l \in \Lambda_k} \av{r_l}$ and we assume that initially no coherences are present. 
\changer{Meanfield will yield qualitatively reasonable predictions unless long-range correlations develop. As we shall show, the transition becomes discontinuous in the quantum regime and is therefore not associated to a divergent correlation length. Our analysis cannot capture the role of spatial dimensionality, but should identify the correct qualitative behavior in sufficiently high dimension, see \cite{buchhold2017}. A more detailed analysis could be achieved exploiting cluster mean-field methods \cite{jin2016cluster}.}
%
%
%
%

In Fig.~\ref{fig:3}(a) we show the stationary state density $N_\mathrm{SS}$ resulting from Eqs.~\eqref{eq:quantum_mf}. A striking difference with respect to the classical case is the appearance of an oscillating behavior as a function of the driving parameter $\Omega/ \kappa$. From our numerical analysis, it appears that the peaks become sharper as the size of the system is increased, with their positions remaining approximately fixed. This suggests that in the thermodynamic limit a sequence of discontinuous jumps will form at fixed values $\wt{\Omega}_j / \kappa$, $(j= 1,2, \ldots)$.
\changer{Note that a standard numerical analysis of the problem (exact diagonalization, quantum-jump Monte Carlo) is forbiddingly complex, due to the exponential scaling of the state space dimension with the number of atoms.}

These jumps feature an intriguing dynamic counterpart: in the quantum case more than one outbreak can occur. As shown in Fig.~\ref{fig:3}(b), if $\Omega / \kappa$ is too small (first column), no outbreak takes place, as in the classical case. For $\Omega \gtrsim {\wt \Omega}_1\approx 0.83 \kappa$, instead, a single outbreak leaves behind an approximately uniform density of immunes (second column). Increasing $\Omega / \kappa$ further, this residual density decreases until, at a second threshold value ${\wt \Omega}_2\approx 1.643 \kappa$, a second outbreak is generated, which causes the final density $N_{\rm{SS}}$ to jump to a higher value. Every new jump in $N_{\rm{SS}}$ appears to be associated to a new outbreak in the dynamics.

This repeating pattern allows us, by analyzing the first jump, to make predictions on the subsequent ones. To this end, we make two simplifying assumptions compatible with the numerically observed behavior: (I) every outbreak leaves behind a uniform density of immunes and (II) stems from the center of the lattice. We focus now on a neighbourhood of the center after the $(j-1)$-th outbreak, at some time $t_j$. By (I) there is an immune density $N_k (t_j) = N^{(j)}>0$; by (II), $R_k$, $G_k$ and $\Sigma_k$ at time $t_j$ correspond to their initial conditions rescaled by $(1-N^{(j)})$. Therefore, the facilitation rate is bounded by $\Omega \overline{R}_k \leq 4 \Omega (1 - N^{(j)})$; as a first approximation, the last factor can be reabsorbed by $\Omega \to \Omega^{(j)} = \Omega (1 - N^{(j)}) $. In other words, the process after $t_j$ proceeds like at $t=0$ (no immunes present), but with a modified frequency $\Omega^{(j)} < \Omega$.

Hence, if $\Omega^{(j)} < \wt{\Omega}_1$ the process stops at $N_{\rm{SS}} = N^{(j)}$, up to subextensive additions,
meaning that $\Omega < \wt{\Omega}_{j}$. If $\wt{\Omega}_1 < \Omega^{(j)} < \wt{\Omega}_2$, instead, a $j$-th outbreak will be produced, but a $(j+1)$-th will not take place, corresponding to $\wt{\Omega}_j < \Omega < \wt{\Omega}_{j+1}$, and so forth. 
All the processes sharing the same reference frequency $\Omega^{(j)}$ are thus equivalent to the same reference process occurring in the absence of immunes and their stationary points will lie on a curve $\Omega (1 - N_{\rm{SS}}) =  const$. The extremal curves in this set, passing through the top and bottom of the first jump, are displayed in Fig.~\ref{fig:3}(a) and bound well the data. With an ansatz $N_{\rm{SS}}\propto 1/\Omega$ for the decrease between jumps, one can formulate a more detailed prediction out of the same considerations, represented by the blue line (see \cite{SM} for details).


\emph{Conclusions ---} We have analyzed a simple model for epidemic spreading in an open quantum system, which has been inspired by recent experimental work \cite{Helmrich2016}, and investigated its connection with the so-called general epidemic process \cite{Grassberger1983}. In the presence of strong dephasing, the process has a direct relation to the GEP, displaying a continuos transition in the same universality class. In the quantum limit, instead, an intriguingly different physics emerges featuring a sequence of discontinuous jumps. This susprising behavior warrants further theoretical and experimental investigation.


\begin{acknowledgments}
The research leading to these results has received funding from the European Research Council under the European Union's Seventh Framework Programme (FP/2007-2013) / ERC Grant Agreement No. 335266 (ESCQUMA), the EPSRC Grant No. EP/M014266/1 and the H2020-FETPROACT-2014 Grant No. 640378 (RYSQ). RG acknowledges the funding received from the European
Union's Horizon 2020 research and innovation programme under the Marie Sklodowska-Curie grant agreement No. 703683. We are also grateful for access to the University of Nottingham High Performance Computing Facility.
\end{acknowledgments}

\bibliographystyle{apsrev4-1}
%

\onecolumngrid
\newpage

\renewcommand\thesection{S\arabic{section}}
\renewcommand\theequation{S\arabic{equation}}
\renewcommand\thefigure{S\arabic{figure}}
\setcounter{equation}{0}
\setcounter{figure}{0}

\begin{center}
{\Large{Supplemental Material: Epidemic dynamics in open quantum spin systems}}
\end{center}

\section{Derivation of the equations of motion in the open quantum system formalism}
We consider an arrangement of quantum 3-level systems on a square lattice. We call the three states $|g\rangle_k$, $|r\rangle_k$ and $|n\rangle_k$, which correspond to the healthy, infected and immune states discussed in the main text, and $k$ denotes the lattice index. For later convenience, we introduce the one-site operators
\be
g_k = \ket{g}_k \bra{g} \!\!\!\!\!\comma\!\!\!\!\! r_k = \ket{r}_k \bra{r} \!\!\!\!\!\comma\!\!\!\!\! n_k  = \ket{n}_k \bra{n} \!\!\!\!\!\comma\!\!\!\!\! \sigma_k^+ = \ket{r}_k \bra{g}  \!\!\!\!\!\comma\!\!\!\!\! \sigma_k^- = \ket{g}_k \bra{r} \!\!\!\!\!\comma\!\!\!\!\! \sigma_k^x = \sigma_k^+ + \sigma_k^- \!\!\!\mand\!\!\! \sigma_k^y = -i \lt \sigma_k^+ - \sigma_k^- \rt.
\ee
As in the main text, we shall also denote by $\Lambda_k$ the set of all lattice indices which are nearest neighbors with site $k$.

The coherent part of the dynamics is generated by a Hamiltonian 
\be
	H = \Omega\sum_k \Pi_k \sigma_k^x,
\ee
where
\be
	\Pi_k  = \sum_{l\in\Lambda_k} r_{l} \prod_{m \in \Lambda_k\backslash \set{l}}  (g_m + n_m)  = \sum_{l\in\Lambda_k} r_{l}   \prod_{m \in \Lambda_k \backslash \set{l}}  (\id - r_m)
\ee
is the projector over all possible configurations in which a single neighbor of site $k$ is in state $\ket{r}$ (infected) and all the others are in superpositions of $\ket{g}$ (healthy) and $\ket{n}$ (immune). 

The dissipative part is instead given in terms of jump operators 
\be
	J^{(\rm{imm})}_k = \sqrt{\kappa} \ket{n}_k \bra{r} \mand J^{(\rm{deph})}_k =  \sqrt{\gamma} \ket{r}_k \bra{r},
\ee
which describe immunization ($\ket{r} \to \ket{n}$) and dephasing (decay of coherence between states $\ket{r}$ and the remaining ones) at site $k$.

The state $\rho$ of the system evolves under the Lindblad equation
\be
	\partial_t \rho = - i\comm{H}{\rho} + \sum_k \lqq \mal{L}(J^{(\rm{imm})}_k) + \mal{L}(J^{(\rm{deph})}_k)   \rqq \rho,
\ee
where
\be
	\mal{L}(J) \rho = J \rho J^\dag - \ha \acomm{J^\dag J}{\rho}.
\ee

\subsection{Effectively classical equations of motion in the large dephasing limit}

We consider here the large dephasing limit $\gamma \gg \Omega$. This condition induces a separation between the timescales on which coherence is produced ($\propto \Omega^{-1}$) and destroyed ($\propto \gamma^{-1}$) and allows to perform an adiabatic elimination of the coherent terms of the density matrix $\rho$ (see Refs.~\cite{Degenfeld14, marcuzzi2014} and the projective Nakajima-Zwanzig method described in \cite{Breuer2002} and references therein). In simpler terms, if the density matrix is written in the basis spanned by the ``classical'' states $\otimes_k\ket{g,r,n}_k$ this allows to write an effective reduced master equation involving just the diagonal part $\mu$, whose elements are simply $\mu_{ij} = \delta_{ij} \rho_{ii}$. This effective equation (see, e.g., \cite{PRL-KinC, Lesanovsky_2014, Everest_2016}) reads, to leading order in $\Omega / \gamma$,
\begin{align}
  \partial_t\mu=\sum_k \left[\alpha \left(\mathcal{L}(\Pi_k \sigma^+_k)+\mathcal{L}(\Pi_k\sigma^-_k)\right)+\kappa\mathcal{L}\left(\left|n\right>_k\!\left<r\right|\right)\right]\mu,
  \label{eq:s_dyn}
\end{align}
with $\alpha  = 4\Omega^2 / \gamma$. This equation is equal to Eq.~\eqref{eq:stochastic_dynamics} in the main text, since $\Pi_k$ is a diagonal projector which only acts on the neighbors of $k$ and therefore $\Pi_k^2 = \Pi_k$ and it commutes with all $k$-th-site-local operators (such as $\sigma_k^{\pm}$) and $\mu$. This is a classical master equation for the classical probabilities $\mu_{ii}$ and describes a stochastic process with three main mechanisms, which we are going to refer to, in the order in which they appear above, as ``classical infection'', ``classical facilitation down'' and ``immunization''. The second process is absent in the general epidemic process and corresponds to the possibility for an infected site to heal a nearby infected site without immunising it at a rate $\alpha$. The corresponding evolution in the Heisenberg picture for a quantity $Q$ (corresponding to a diagonal operator in the classical basis) reads now
\be
	\partial_t Q = \sum_k \lqq  \alpha  \lt \mal{L}^\ast ( \Pi_k \sigma_k^+) + \mal{L}^\ast ( \Pi_k \sigma_k^-) \rt + \kappa \mal{L}^\ast (\ket{n}_k \bra{r})  \rqq Q,
\ee
where $\mal{L}^\ast$ is the adjoint dissipator
\be
	\mal{L}^\ast (J) Q = J^\dag Q J - \ha \acomm{J^\dag J}{Q}.
	\label{eq:adj}
\ee
In the following, since higher-order terms in the immune density are not expected to modify the physics of the transitions studied in the main text, we shall replace $\Pi_k$ by a softer constraint, which makes the rate proportional to the number of infected neighbors a site has. The corresponding simplified equation reads
\be
	\partial_t Q = \sum_k \lqq  \alpha  \sum_{l \in \Lambda_k} \lt \mal{L}^\ast ( r_l \sigma_k^+) + \mal{L}^\ast ( r_l \sigma_k^-) \rt + \kappa \mal{L}^\ast (\ket{n}_k \bra{r})  \rqq Q.
\ee
Since this is a linear equation, one can study it term by term. The relevant evolution equations for the one-site operators are

\begin{itemize}
  \item classical infection
  \begin{eqnarray}
  \partial_t r_m &=& \alpha \sum_k \sum_{l \epsilon \Lambda_k} \lqq  r_l \sigma^-_k r_m \sigma^+_k r_l - \frac{1}{2} r_l \sigma^-_k \sigma^+_k r_l r_m- \frac{1}{2} r_m r_l \sigma^-_k \sigma^+_k r_l \rqq \nonumber \\
  &=&  \alpha \sum_k \sum_{l \epsilon \Lambda_k} \lqq r_l \sigma^-_k r_m \sigma^+_k r_l - r_m r_l g_k \rqq =  \alpha  \sum_{l \epsilon \Lambda_m} r_l g_m, \nonumber \\
  \partial_t n_m &=& 0
  \end{eqnarray}
\item classical facilitation down
\begin{eqnarray}
  \partial_t r_m &=& \alpha \sum_k \sum_{l \epsilon \Lambda_k} \lqq  r_l \sigma^+_k r_m \sigma^-_k r_l - \frac{1}{2} r_l \sigma^+_k \sigma^-_k r_l r_m- \frac{1}{2} r_m r_l \sigma^+_k \sigma^-_k r_l  \rqq \nonumber  \\
  &=&  \alpha \sum_k  \sum_{l \epsilon \Lambda_k} \lqq  r_l \sigma^+_k r_m \sigma^-_k r_l - r_m r_l r_k \rqq =-\alpha \sum_{l \epsilon \Lambda_m} r_l r_m, \nonumber \\
  \partial_t n_m &=& 0
  \label{eq:clfd}
\end{eqnarray}
\item immunization
\begin{eqnarray}
  \partial_t r_m &=& \kappa \sum_k \lqq |r\rangle_k\langle n| r_m |n\rangle_k\langle r| - \frac{1}{2} r_k r_m- \frac{1}{2} r_m r_k \rqq=-\kappa r_m,\nonumber \\
  \partial_t n_m &=& \partial_t |n\rangle_m\langle n| = \kappa \sum_k \lqq |r\rangle_k\langle n| n_m |n\rangle_k\langle r| - \frac{1}{2} r_k n_m- \frac{1}{2} n_m r_k \rqq = \kappa r_m.
\end{eqnarray}
\end{itemize}
All the remaining terms can be obtained via the local probability conservation $r_m + g_m + n_m = 1$ (the probability of a site being in \emph{any} state is $1$). Employing the shorthand $\overline{r}_m=\sum_{l \epsilon \Lambda_m} r_l$ we arrive at
\begin{eqnarray}
  \partial_t r_m&=& \alpha \overline{r}_m (g_m-r_m) -\kappa r_m,\\
  \partial_t g_m&=& - \alpha \overline{r}_m (g_m-r_m),\\
  \partial_t n_m&=&\kappa r_m.
\end{eqnarray}
We now introduce the notation $N_m = \av{n_m}$, $R_m = \av{r_m}$, $G_m = \av{g_m}$ and $\overline{R}_m = \av{\overline{r}_m}$ for the expectation value of these observables and perform a mean-field approximation. This means we factorize all correlation functions acting on different sites, i.e., $\av{Q_k Q_m} \to \av{Q_k} \av{Q_m}$ for $k \neq m$ and for any one-site-local quantity $Q$. The equations then read
\begin{eqnarray}
  \partial_t R_m&=& \alpha \overline{R}_m (G_m-R_m) -\kappa R_m,\\
  \partial_t G_m&=& - \alpha \overline{R}_m (G_m-R_m),\\
  \partial_t N_m&=&\kappa R_m.
\end{eqnarray}
If we additionally assume translational invariance and introduce the notation $N = \sum_k\av{n_k}/L$, $R = \sum_k\av{r_k}/L$, $G = \sum_k\av{g_k}/L$ and the coordination number $z$ (number of nearest neighbors per site, $4$ for a square lattice) we arrive at
\begin{eqnarray}
  \partial_t R&=& z\alpha R (G-R) -\kappa R, \label{eq:Rcl}\\
  \partial_t G&=& - z\alpha R(G - R) \label{eq:Gcl},\\
  \partial_t N&=&\kappa R \label{eq:Ncl}.
\end{eqnarray}

\subsection{Equations of motion in the quantum case}

For $\gamma = 0$, the derivation of the equations of motion for the one-site observables proceeds along the same lines, with the adjoint Lindblad equation now reading
\be
	\partial_t O = i \comm{H}{O} + \kappa \sum_k \mal{L}^\ast(\ket{n}_k \bra{r}) O 
\ee
for any observable $O$ and $\mal{L}^\ast$ as in \eqref{eq:adj}. For simplicity, we consider here the softer constraint $\Pi_k \to \overline{r}_k$ as well. We can now write again the various contributions term by term, which read
\begin{itemize}
\item immunization
\begin{eqnarray}
  \partial_t r_m &=& \kappa \sum_k \lqq |r\rangle_k\langle n| r_m |n\rangle_k\langle r| - \frac{1}{2} r_k r_m- \frac{1}{2} r_m r_k \rqq=-\kappa r_m
\end{eqnarray}
\begin{eqnarray}
  \partial_t n_m &=& \partial_t |n\rangle_m\langle n| = \kappa \sum_k \lqq |r\rangle_k\langle n| n_m |n\rangle_k\langle r| - \frac{1}{2} r_k n_m- \frac{1}{2} n_m r_k \rqq = \kappa r_m
\end{eqnarray}
\begin{eqnarray}
  \partial_t \sigma^-_m &=& \kappa \sum_k \lqq|r\rangle_k\langle n|\sigma^-_m |n\rangle_k\langle r| - \frac{1}{2} r_k \sigma^-_m- \frac{1}{2} \sigma^-_m r_k \rqq=-\frac{\kappa}{2} \sigma^-_m
\end{eqnarray}
\item quantum facilitation
\begin{eqnarray}
  \partial_t r_m &=& i\Omega \sum_k \sum_{l \epsilon \Lambda_k} [r_l\sigma_k^x,r_m]=i\Omega \sum_k \sum_{l \epsilon \Lambda_k} r_l[\sigma_k^x,r_m]=\Omega \sum_{l \epsilon \Lambda_m} r_l \sigma^y_m
\end{eqnarray}
\begin{eqnarray}
  \partial_t \sigma^-_m &=& i\Omega \sum_k \sum_{l \epsilon \Lambda_k} [r_l\sigma_k^x,\sigma^-_m]=i\Omega \sum_k \sum_{l \epsilon \Lambda_k} r_l[\sigma_k^x,\sigma^-_m]+[r_l,\sigma^-_m]\sigma_k^x \nonumber \\
  &=&i\Omega \sum_k \sum_{l \epsilon \Lambda_k} \lqq r_l(r_m-g_m)\delta_{km}-\sigma^-_m\sigma_k^x\delta_{lm} \rqq \nonumber \\
  &=&i\Omega \sum_{l \epsilon \Lambda_m} [r_l(r_m-g_m)-\sigma_l^x\sigma^-_m]
\end{eqnarray}
\end{itemize}
The equations for $\sigma^+_m$ can be derived by taking the hermitian conjugate of the equations for $\sigma^-_m$, and those for $\sigma_m^{x/y}$ simply by linear combination according to the definitions. Combining all terms and defining $\overline{s}^x_m = \sum_{l \in \Lambda_m} \sigma^x_l$, in this case the equations read
\begin{eqnarray}
  \partial_t r_m&=&\Omega \overline{r}_m \sigma^y_m  -\kappa r_m,\\
  \partial_t g_m&=&-\Omega \overline{r}_m \sigma^y_m ,\\
  \partial_t n_m&=&\kappa r_m,\\
  \partial_t \sigma^y_m&=&-2\Omega \overline{r}_m(r_m - g_m)+ \Omega \overline{s}^x_m \sigma^x_m - \frac{\kappa}{2} \sigma^y_m,\\
  \partial_t \sigma^x_m&=&-\Omega \overline{s}^x_m \sigma^y_m -\frac{\kappa}{2}\sigma^x_m.
  \end{eqnarray}
The mean-field approximation yields in this case the equations
\begin{eqnarray}
  \partial_t R_m&=&\Omega \overline{R}_m \Sigma^y_m  -\kappa R_m,\\
  \partial_t G_m&=&-\Omega \overline{R}_m \Sigma^y_m ,\\
  \partial_t N_m&=&\kappa R_m,\\
  \partial_t \Sigma^y_m&=&-2\Omega \overline{R}_m(R_m - G_m)+ \Omega \overline{S}^x_m \Sigma^x_m - \frac{\kappa}{2} \Sigma^y_m,\\
  \partial_t \Sigma^x_m&=&-\Omega \overline{S}^x_m \Sigma^y_m -\frac{\kappa}{2}\Sigma^x_m,
  \end{eqnarray}
where we additionally defined $\overline{S}_m^x = \av{\overline{s}^x_m}$ and $\Sigma^{x/y}_m = \av{\sigma^{x/y}_m}$. We see now that, if $\Sigma_m^x = 0~\forall m$, then $\partial_t \Sigma_m^x = 0~\forall m$ and the $x$ component of the coherences remains null in the mean-field dynamics. Restricting to initial conditions with vanishing coherences, one can thus neglect the last equation altogether and work with the reduced set
 \begin{eqnarray}
  \partial_t R_m&=&\Omega \overline{R}_m \Sigma^y_m  -\kappa R_m, \label{eq:Q1}\\
  \partial_t G_m&=&-\Omega \overline{R}_m \Sigma^y_m ,\\
  \partial_t N_m&=&\kappa R_m,\\
  \partial_t \Sigma^y_m&=&-2\Omega \overline{R}_m(R_m - G_m) - \frac{\kappa}{2} \Sigma^y_m \label{eq:Q4}, 
\end{eqnarray}
and make the substitution $\Sigma^y_m \to \Sigma_m$, which reproduces Eqs.~\eqref{eq:quantum_mf} in the main text.
Introducing the notation $\Sigma = \sum_m \av{\sigma_m}/L$, the uniform case reads now 
\begin{eqnarray}
  \partial_t R&=&\Omega zR \Sigma^y -\kappa R \label{eq:Rq}\\
  \partial_t G&=&-\Omega zR \Sigma^y \label{eq:Gq}\\
  \partial_t N&=&\kappa R \label{eq:Nq}\\
  \partial_t \Sigma & = & -2\Omega zR(R-G) -\frac{\kappa}{2} \Sigma. \label{eq:Sq}
\end{eqnarray}

\section{Properties of the homogeneous mean-field equations}

\subsection{Classical equation limit}
In the large dephasing limit $\gamma \gg \Omega$ Eqs.~\eqref{eq:Rcl} and \eqref{eq:Gcl} close among themselves. The fact that $R = 0$ is sufficient to annihilate both derivatives shows that, in principle, any value of $G$ is admitted in the stationary state, which reflects the fact that any configuration with only healthy and immune sites is absorbing for the stochastic process. The specific value of $G$ that the dynamics asymptotically reaches is then established by the initial condition, which hints at the presence of a constant of motion $F(G,R,N)$ ($\partial_t F(G,R,N) = 0$), apart from the trivial one $G + R + N$, which carries the corresponding information. This is analogous to the considerations found in \cite{Grassberger1983, Janssen1985} for the deterministic equations. In our case, the equation for the difference
\be
	\partial_t (R - G) = -2z\alpha R\left(R - G + \frac{\kappa}{2z\alpha}\right)
\ee
can be recast in the form
\be
	\partial_t \log \lt R - G + \frac{\kappa}{2z\alpha}  \rt = -2z\alpha R =  -\frac{2z\alpha}{\kappa} \partial_t N, 
\ee
which identifies
\be
	F(G,R,N) = \log \lt R - G + \frac{\kappa}{2z\alpha}  \rt + \frac{2z\alpha}{\kappa} N
\ee
and can be implicitly integrated to yield
\be
	R - G = -\frac{\kappa}{2z\alpha} + \lqq  R_0 - G_0 + \frac{\kappa}{2z\alpha} \rqq \rme{-\frac{2z\alpha}{\kappa} (N - N_0)},
\ee
where $R_0$, $G_0$ and $N_0$ denote the initial conditions. Exploiting the conservation of probability $R + G + N = 1$, we can also write 
\be
	2R + N  - 1 = -\frac{\kappa}{2z\alpha} + \lqq  R_0 - G_0 + \frac{\kappa}{2z\alpha} \rqq \rme{-\frac{2z\alpha}{\kappa} (N - N_0)}
\ee
and, since $R = \partial_t N / \kappa$, after some rearrangements,
\be
	\frac{2}{\kappa} \partial_t N = 1 - N -\frac{\kappa}{2z\alpha} + \lqq  R_0 - G_0 + \frac{\kappa}{2z\alpha} \rqq \rme{-\frac{2z\alpha}{\kappa} (N - N_0)}.
\ee
Abbreviating $\xi = 2z\alpha/\kappa$ and rescaling time by $\tau = \kappa t$ we find the more compact expression
\be
	\partial_\tau N = \ha \lqq 1-N - \frac{1}{\xi} + \lt  R_0 - G_0 + \frac{1}{\xi} \rt\rme{-\xi (N - N_0)}      \rqq.
\ee
Note that this single equation for $N$ is not generally equivalent to Eqs.~\eqref{eq:Rcl}, \eqref{eq:Gcl}, \eqref{eq:Ncl}: it only yields the same result as long as \emph{the same} initial conditions $R_0$ and $G_0$ for $R$ and $G$ are taken.
Finally, setting $N_0 = 0$, $R_0 = \epsilon$, $G_0 = 1-\epsilon$ (initially a fraction $\epsilon$ of sites is in state $|r\rangle$, while the rest are in state $|g\rangle$), one finds
\be
	\partial_\tau N = \ha \lqq  1 - N - \frac{1}{\xi} + \left(2\epsilon -1+ \frac{1}{\xi}\right)\, \rme{-\xi N}   \rqq = \ha \lqq - N + \left(1 - \frac{1}{\xi}\right)\lt 1 - \rme{-\xi N} \rt + 2\epsilon\, \rme{-\xi N}  \rqq.
\label{eqforN}
\ee
In order to study the stationary properties (at initial conditions fixed as above), we now consider the long-time limit and set $\partial_\tau N = 0$. The initial condition we assume in the main text is of a single infected site placed at the centre of the system at $ t = 0$, corresponding to a vanishing initial density ($\epsilon \to 0^+$) in the thermodynamic limit. We thus set $\epsilon = 0$ as well in \eqref{eqforN}. Note that the limits $t\to\infty$ and $\epsilon \to 0^+$ do not commute: starting the dynamics with $R_0 = 0$ will yield the trivial solution $R = 0$, $G = G_0$, $N = N_0$. The stationary equation now reads
\be
	\frac{N_\mathrm{SS}}{1-\frac{1}{\xi}} =  1 - \rme{-\xi N_\mathrm{SS}} ,
	\label{eq:critp}
\ee
where $N_\mathrm{SS}$ stands for the density of immunes in the stationary state. Since $\xi \geq 0$, the function on the r.h.s.~is concave, whereas the one on the l.h.s.~is just a line. Hence, at any value of $\xi$ there can be at most two intersections. One is always present at $N_\mathrm{SS} = 0$. Evaluating the derivatives in $0$, it is not difficult to show that the line is tangent to the curve for $\xi = 2$. For $\xi > 2$ a second point of intersection appears at some value $0 < N_\mathrm{SS} < 1$, while the solution $N_\mathrm{SS} = 0$ becomes unstable. For $1 < \xi < 2$ this second point lies at negative values and is thus unphysical. For $\xi \leq 1$ there is only the vanishing solution. Hence, $\xi_c = 2$ marks a critical point where the system switches from a non-percolating phase ($\xi < 2$) where the process always stops at some finite time, and a percolating phase ($\xi > 2$) where the process has a finite probability of surviving for arbitrary times.

For $\xi > 2$ but close to the critical point, the stationary density of immune sites $N$ remains small and one can Taylor expand Eq.~\eqref{eq:critp} to find
\be
	\frac{N_\mathrm{SS}}{1-\frac{1}{\xi}} \approx \xi N_\mathrm{SS}- \ha \xi^2 N_\mathrm{SS}^2 \quad \Rightarrow \quad N_\mathrm{SS} \sim  \xi - 2 .
\ee
Thus, the mean-field prediction for the order parameter critical exponent is $\beta_{MF} = 1$. 

\begin{figure}[h]
	\includegraphics[width=0.5\columnwidth]{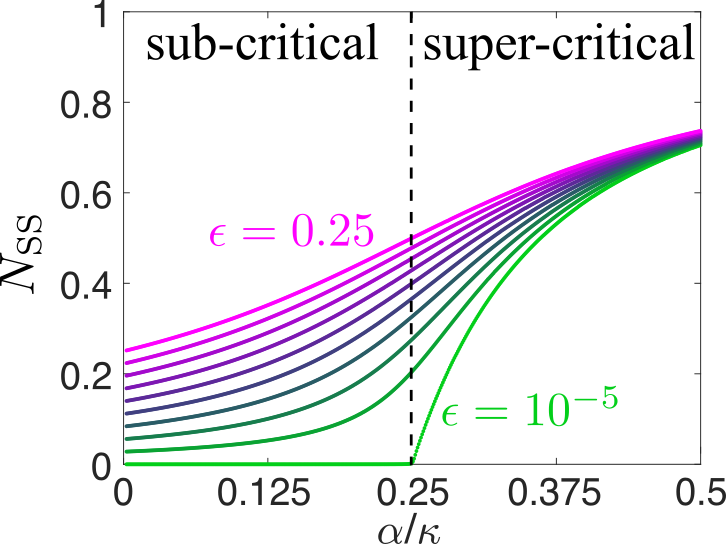}
\caption{Dependence of the stationary state value of $r$ on the initial conditions (classical limit $\Omega=0$). We have parameterized the initial conditions as $G(0)=1-\epsilon$, $R(0)=\epsilon$ and $N(0)=0$ and propagated the system up to $\kappa t_\mathrm{final}=500$. The different curves are for different values of $\epsilon$. Note that the value of $N_\mathrm{SS}$ at $\alpha=0$ is by construction equal to $\epsilon$. The curves show that in the limit $\epsilon\rightarrow 0^+$ the typical behaviour of a continuous phase transition is displayed. This plot appears as Fig. 2(a) in the main text.}
\label{fig:init_conditions_classical}
\end{figure}

In Figure \ref{fig:init_conditions_classical}, we show the meanfield solution \eqref{eqforN} for different values of $\epsilon$, where one can see a crossover between the sub- and super-critical phases that approaches a continuous phase transition as $\epsilon \to 0^+$. These solutions have been found numerically, assuming the underlying lattice is square, and therefore $z=4$, which leads to $(\alpha/\kappa)_c = \frac{\xi_c}{8} = \frac{1}{4}$.

\vspace{0.5cm}

\subsection{Quantum regime}
\begin{figure}[h]
	\includegraphics[width=0.5\columnwidth]{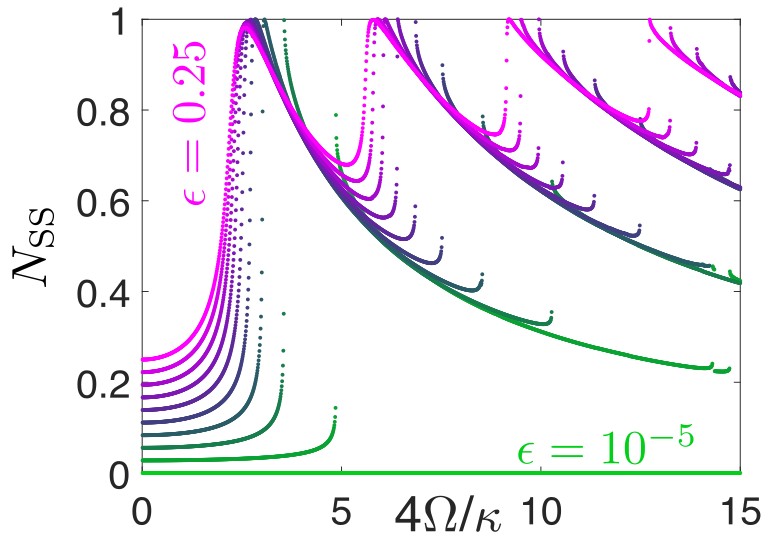}
\caption{Dependence of the stationary state value of $N$ on the initial conditions (quantum limit $\alpha=0$). We have parameterized the initial conditions as $G_0=1-\epsilon$, $R_0=\epsilon$ and $N_0=0$ and propagated the system up to $\kappa t_\mathrm{final}=500$. Different curves correspond to different values of $\epsilon$. Note that the value of $N_\mathrm{ss}$ at $\Omega=0$ is by construction equal to $\epsilon$..}
\label{fig:init_conditions_quantum}
\end{figure}

In the quantum regime $(\gamma=0)$ the equations of motion \eqref{eq:Rq}-\eqref{eq:Sq} are such that $R = \Sigma = 0$ makes all derivatives vanish. As in the classical case, all states with no coherence nor component over the states $\ket{r}_k$ are valid stationary states of the system. The values reached by $G$ and $N$ in the long-time limit are hence once again dependent on the initial conditions. In this case, however, we have been unable to identify a constant of motion (apart from the trivial one $N + R + G = 1$) to connect the initial and final values.
We thus resort to a numerical analysis. For the initial conditions $G(0)=1-\epsilon$, $R(0)=\epsilon$ and $N(0)=\Sigma(0)=0$, and again taking $z=4$, we find the stationary state density $N_\mathrm{SS}$ displayed in Fig.~\ref{fig:init_conditions_quantum}. A closer inspection of these curves suggests that, despite the abrupt increases that look like discontinuities due to the numerical resolution, they do not feature any discontinuous behavior.

As pointed out in the main text, a clear difference with respect to the classical case is the appearance of an oscillating behavior of $N_{\rm{SS}}$ in $\Omega/ \kappa$. For the uniform mean-field equations, the oscillation peaks tend to shift to larger values of $\Omega / \kappa$ as the initial density of infected sites, $R(0) = \epsilon$, is decreased. For $\epsilon=10^{-5}$ the system remains in a state with $N_\mathrm{SS} \approx 0$ over the whole parameter range shown. We have verified that, in the parameter range shown, the position of the first jump scales as $\wt{\Omega}_1 \sim \kappa/(\sqrt{\epsilon})$ for small $\epsilon$. Extrapolating this behaviour to $\epsilon \to 0^+$ indicates that, in contrast to the classical case, the homogeneous meanfield equations do not lead to epidemic spreading for any finite $\Omega$ as long as the initial infected density is vanishingly small ($\epsilon \to 0^+$).

The reason for the discrepancy between the predictions obtained from the homogeneous meanfield equations and the space-dependent meanfield dynamics explored in the main text seems to be related to the ambiguity existing in the definition of the initial condition. While meant to reproduce the situation of a single infected site among $L$, in fact, the uniform mean-field equations only admit uniform initial conditions. Conceptually, the closest one to reality is one where $R(0) = 1/L$ while $G(0) = 1-1/L$. In the classical case, this can be interpreted as a uniform probability density of finding a single particle anywhere in the system. In the quantum case, instead, another interpretation in terms of quantum superpositions is possible: one could have, in fact, every site in state $\sqrt{L^{-1}}\, |r\rangle + \sqrt{1 - L^{-1}}\,|g\rangle$. It is very much possible that, due to the different structure of the equations, the quantum process is unable to give rise to a self-sustaining infection in the limit $L \to \infty$ if $\Omega$ is kept finite. The numerical solution of the inhomogeneous equations \eqref{eq:Q1}-\eqref{eq:Q4} shows a very different behaviour (as discussed in the main text) when instead we set the real initial condition of having a single site in state $\ket{r}$ in the center and all the others in state $\ket{g}$.

\section{Critical behavior of the epidemic dynamics in the classical limit}
In order to determine the universality class of the classical stochastic dynamics given by Eqs.~\eqref{eq:pik}-\eqref{eq:stochastic_dynamics} in the main text, we have performed a finite size scaling analysis for both the 
full constraint $\Pi_k = \sum_{l\in \Lambda_k}r_l\prod_{m \in \Lambda_k\backslash \set{l}}(g_m+n_m)$ and its linear approximation ($\Pi_k\approx \sum_{l\in \Lambda_k}r_l$). Performing the same finite size scaling analysis of Ref.~\cite{argolo2011finite} we have first derived the critical value of the control parameter $\alpha_c (L)$ (setting $\kappa=1$ for simplicity) by locating the maximum of the numerical derivative of the order parameter $N_{\text{SS}}$ with respect to $\alpha$ or, in other words, the inflection point of the curve, where it switches from being convex to concave. In the thermodynamic limit $L \to \infty$, this point will approach the actual critical value $\alpha_c$, where a non-analiticity develops, signalling the presence of a second order phase transition. Close to this point, in the so-called critical scaling region, observables such as the stationary density of immunes $N_{\rm{SS}}$ will acquire a scaling form, i.e., under a rescaling of all distances by a factor $b$
\be
	N_{\rm{SS}} \lt \alpha - \alpha_c, L    \rt = b^{\Delta_N} N_{\rm{SS}} \lt b^{-\Delta_\alpha} (\alpha - \alpha_c) , b^2 L  \rt,
\ee
where $\Delta_N$ and $\Delta_\alpha$ are the scaling dimensions of $N_{\rm{SS}}$ and the control parameter $\alpha - \alpha_c$, respectively, and we recall that, the system being two dimensional, $L$ scales like a surface (i.e., $\propto b^2$). By fixing $b = 1/\sqrt{L}$ we can then write
\be 
	N_{\rm{SS}} \lt \alpha - \alpha_c, L \rt = L^{-\frac{\Delta_N}{2}} N_{\rm{SS}} \lt L^{\frac{\Delta_\alpha}{2}} \lt \alpha - \alpha_c \rt , 1 \rt = L^{-\frac{\Delta_N}{2}} f \lt L^{\frac{\Delta_\alpha}{2}} \lt \alpha - \alpha_c \rt \rt
\ee
with $f$ a universal (up to a multiplicative constant) scaling function. Using standard conventions, we relate the scaling dimensions to the static critical exponents $\beta$ (do not confuse this $\beta$ with the one of Eq. (6) in the main text) and $\nu$ (see e.g.~Ref.~\cite{Grassberger1983}) according to
\be
	\Delta_\alpha = \frac{1}{\nu} \quad \quad \Delta_N = \frac{\beta}{\nu}, 
\ee
which therefore implies that 
\be
	L^\frac{\beta}{2\nu} N_{\rm{SS}} = f \lt L^{\frac{\Delta_\alpha}{2}} (\alpha - \alpha_c) \rt,
\ee
i.e., once rescaled by $L^\frac{\beta}{2\nu}$ and plotted as functions of $L^{\frac{\Delta_\alpha}{2}} (\alpha - \alpha_c)$, all datasets obtained at different system sizes should collapse onto the same master curve $f$.

\begin{figure}[h]
	\includegraphics[width=0.5\columnwidth]{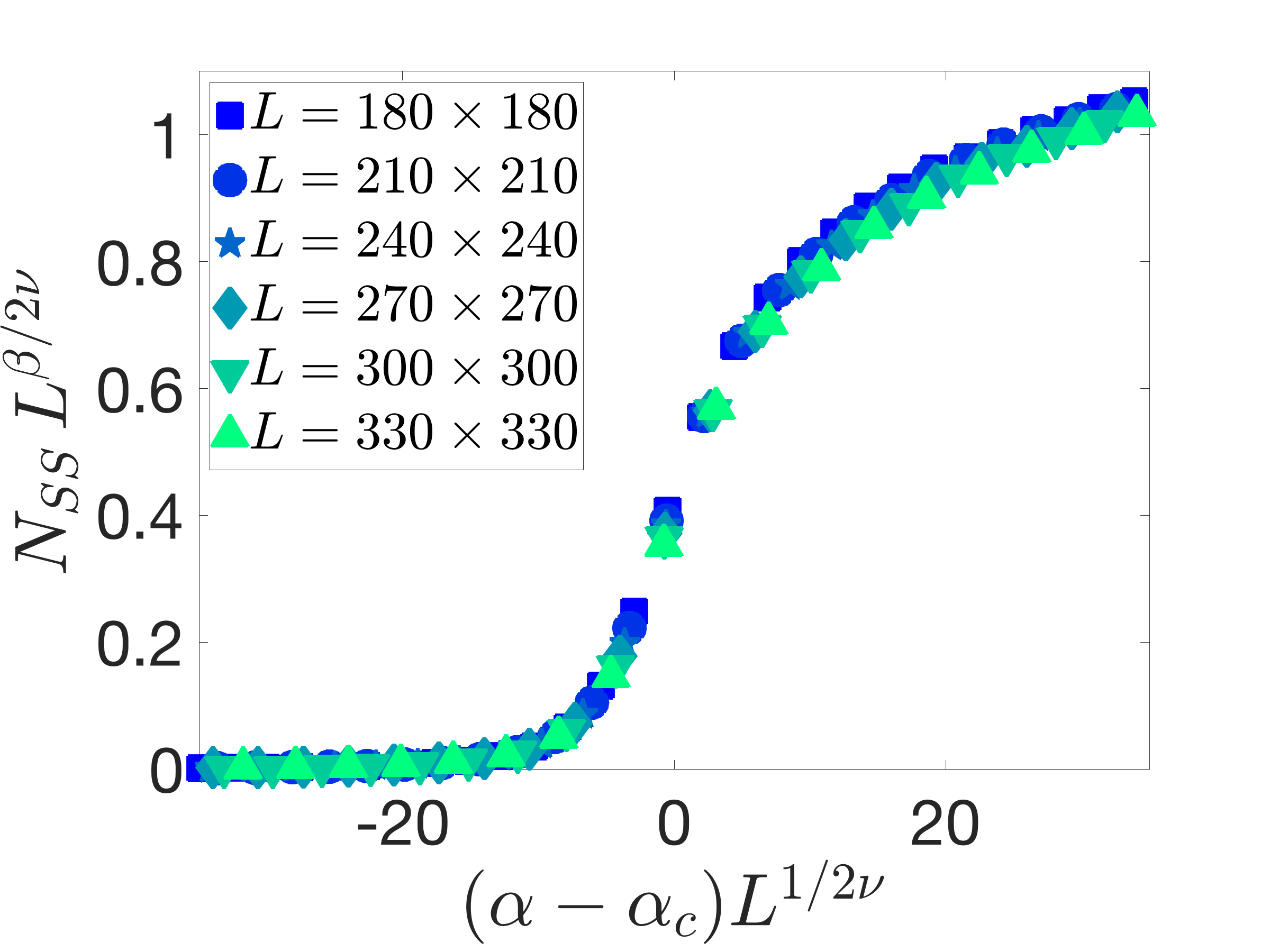}
\caption{Data collapse of the mean density of immune states for the classical stochastic dynamics with the constraint $\Pi_k\approx \sum_{l\in \Lambda_k}r_l$ approximated by the leading linear terms, including curves for six different system sizes, as reported in the legend. The critical exponents $\beta$ and $\nu$ correspond to those of the dynamic isotropic percolation universality class ($\beta=5/36$ and $\nu=4/3$). The critical value $\alpha_c=1.71\pm0.01$ was taken as the inflection point of the curve at larger system size (we recall that we set for simplicity $\kappa=1$).} 
\label{fig:collapse_sum}
\end{figure}

\begin{figure}[h]
	\includegraphics[width=0.5\columnwidth]{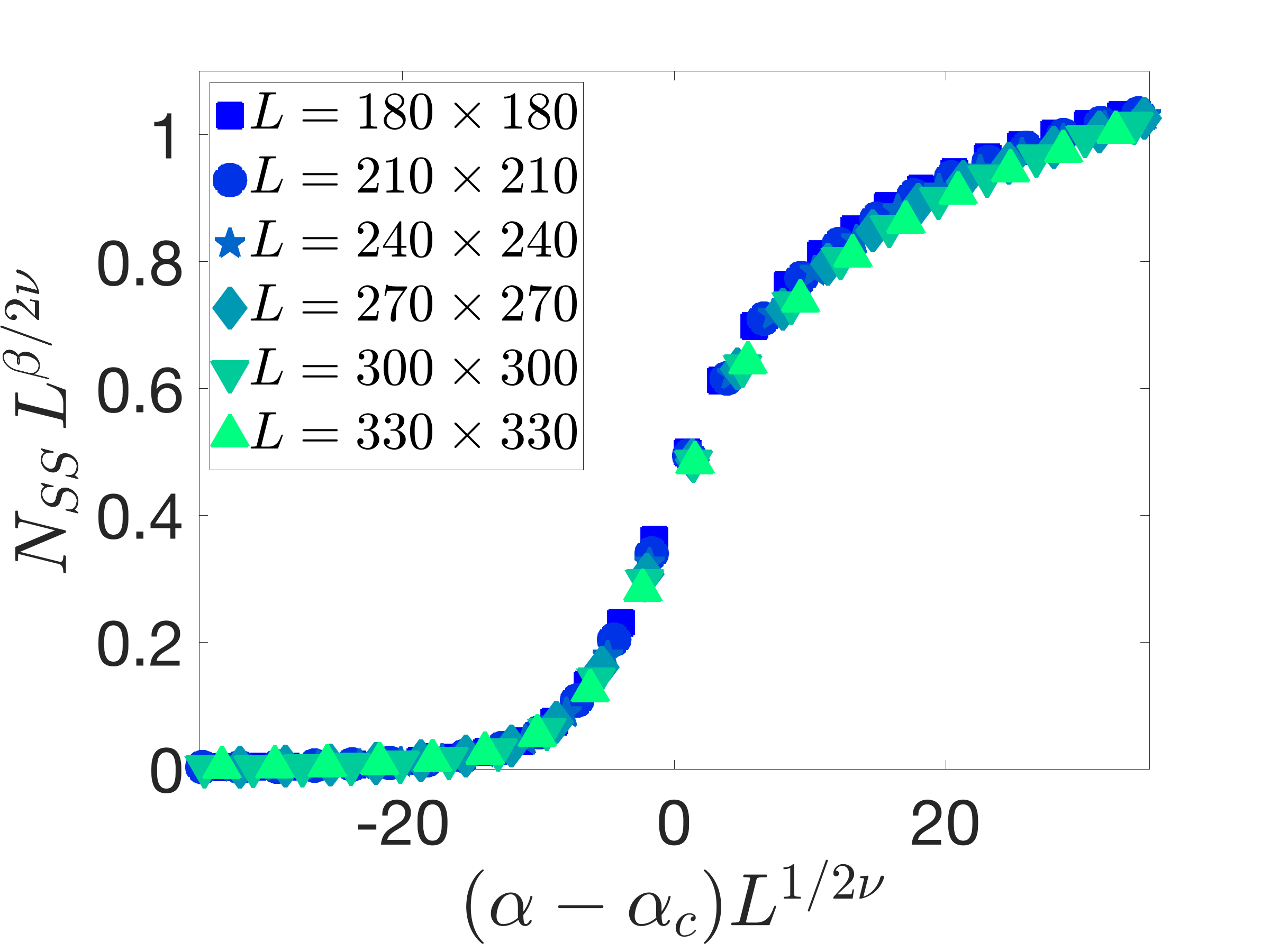}
\caption{Data collapse of the mean density of immune states for the classical stochastic dynamics with the full constraint $\Pi_k = \sum_{l\in \Lambda_k}r_l\prod_{m \in \Lambda_k\backslash \set{l}}(g_m+n_m)$ in Eq.~\eqref{eq:pik} in the main text for the same system sizes explored in Fig.~\ref{fig:collapse_sum} above. The critical exponents have been taken from the tabulated values for dynamic isotropic percolation universality ($\beta=5/36$ and $\nu=4/3$) and $\alpha_c=1.93\pm0.01$ ($\kappa = 1$) has been estimated as the position of the inflection point of the curve at largest system size.} 
\label{fig:collapse_jone}
\end{figure}

In order to verify this, we have used the critical exponents characterizing the general epidemic process, which are those of the dynamic isotropic percolation universality class \cite{Grassberger1983} ($\beta=5/36$ and $\nu=4/3$). The numerical results are displayed in Figs.~\ref{fig:collapse_sum} (GEP constraint $\overline{R}_k$, i.e.~facilitation rate proportional to the number of infected neighbors) and \ref{fig:collapse_jone} (stricter constraint $\Pi_k$, i.e.~facilitation rate only $>0$ when a single infected neighbor is present) for different system sizes, where one can observe a collapse for both types of dynamics, strongly indicating that they belong to the same universality class and that the latter is isotropic percolation for the stationary properties.

\subsection{Irrelevance of higher-order terms in the infected-site density}

We exploit here the results known for the effective field-theoretical description of the GEP \cite{cardy1983field, Janssen1985, cardy1985epidemic} to verify that the new terms generated by (A) the classical facilitation down \eqref{eq:clfd} and by (B) the more stringent constraint $\overline{R}_k \to \Pi_k$ are irrelevant in a renormalization-group sense and thus are not expected to alter the universal properties of the transition. First, let us rewrite the classical uniform mean-field equations \eqref{eq:Rcl}-\eqref{eq:Ncl} as
\begin{eqnarray}
  \partial_t R&=& z\alpha R (G - cR) \mal{F} -\kappa R, \\
  \partial_t G&=& - z\alpha  R  (G - cR) \mal{F} ,\\
  \partial_t N&=&\kappa R ,
\end{eqnarray}
where now we distinguish
\begin{align}
	&\mal{F}^{(GEP)} = 1 ,  & &c^{(GEP)} = 0, \\
	&\mal{F}^{(GEP) + (A)} = 1 , & &c^{(GEP) + (A)} = 1, \\
	&\mal{F}^{(GEP) + (A)+ (B)} = (1-R)^{z-1} , & &c^{(GEP) + (A) + (B)} = 1.
\end{align}
Following Ref.~\cite{Janssen1985}, we start from the deterministic equations and formally integrate the last two to find
\be
\begin{split}
	G(t) &= G_0 \rme{-z \alpha \int_0^t \rmd \tau R(\tau) \mal{F}(R(\tau)) } + zc\alpha \int_0^t \rmd \tau \, \rme{-z\alpha \int_\tau^t \rmd \tau' R(\tau') \mal{F}(R(\tau'))} R^2(\tau) \mal{F}(R(\tau)) \\
	N(t) &= \kappa \int_0^t  \rmd \tau \, R(\tau) .
\end{split}
\ee
The implicit expression for $G(t)$ as a function of $R(t)$ can now be substituted in the first equation. The subsequent step would be to expand the resulting r.h.s.~in powers of the infected density $R$, keeping only the lowest orders. We thus immediately notice that
\be 
	\mal{F}^{(GEP) + (A)+ (B)} = (1-R)^{z-1} = 1 - (z-1) R + \matb{c} z-1 \\ 2 \mate R^2 + \ldots = 1 + O(R)
\ee
is equivalent to $\mal{F}^{(GEP)}$ up to higher orders, which would at most rescale the coupling constants by some amount. In particular, at lowest order
\be
	(G - cR)(1-R)^{z-1} = G - (1+(z-1)G) R + O(R^2),
\ee
which shows that, since $z>1$ and $G \geq 0$, the correction never makes the linear term vanish (i.e., it amounts to an irrelevant rescaling which can be reabsorbed in the coupling itself). For this reason, in the following we only consider $\mal{F}^{(GEP)} = 1$, to avoid purposeless complications. This will also make it relatively easy to distinguish GEP terms (surviving when $c = 0$) from new ones (vanishing for $c = 0$). For later convenience, we also introduce the notation
\be
	\mal{R}_1 (t) = \int_0^t \rmd \tau\, R(\tau) \mand \mal{R}_2 (t) = \int_0^t \rmd \tau\, R(\tau)^2.
\ee
By Taylor expanding in $R$ we thus find
\be
	G(t) = G_0 - z\alpha G_0 \mal{R}_1 (t)  + \lqq  G_0 \frac{\lt z\alpha  \rt^2}{2} \mal{R}_1 (t)^2 + zc\alpha \mal{R}_2 (t)   \rqq + O(R^3).
\ee
One can now re-establish the dependence on spatial coordinates by switching the variables with continuous classical fields $R(t) \to R(\vec{x},t)$, $\mal{R}_i (t) \to \mal{R}_i (\vec{x},t)$ and furthermore introduce stochastic fluctuations in the form of a random Gaussian source $\xi{\vec{x},t}$ which obeys
\be 
	\av{\xi(\vec{x},t)} = 0  \mand   \av{\xi(\vec{x}, t)\, \xi(\vec{y},t') } = K R(\vec{x},t)  \delta(\vec{x} - \vec{y}) \delta(t-t'),
\ee
with $K$ a positive constant and where the proportionality of the variance to the amplitude of $R$ accounts for the fact that all configurations with $R \equiv 0$ are absorbing, and thus fluctuationless --- i.e., whenever the system becomes depleted of infected individuals the stochastic dynamics halts.

The equation of motion for $R$ then takes the form of a Langevin equation
\be
	\partial_t R = (D \nabla^2 + \underbrace{z\alpha G_0 - \kappa}_{-u_2}) R - \underbrace{(z\alpha)^2 G_0}_{u_3} R \mal{R}_1 + \underbrace{G_0 \frac{(z\alpha)^3}{2}}_{u_4} R \mal{R}_1^2 {\textcolor{red}{ -  \underbrace{z \alpha c}_{c_3} R^2 + \underbrace{z c \alpha }_{c_4} R \mal{R}_2 }} + \xi ,
	\label{eq:Langevin}
\ee
where the space and time dependence is intended and only the first (diffusive) term in a derivative expansion is kept with an effective coupling $D$. We have introduced shorthands for the coupling constants, where the $u_i$s belong to the GEP, whereas the $c_i$s are ``new''. For the reader's convenience, the ``non-GEP'' terms have also been highlighted in red color. 

Langevin equations can be generically mapped onto equivalent effective field theories via the so-called Martin-Siggia-Rose-de Dominicis-Janssen (MSRDJ) formalism \cite{Tauber-book}. Here we do not review it, but simply exploit the results found in Ref.~\cite{Janssen1985} to determine the engineering scaling dimension of the new terms. To this end, we should mention that the construction of the effective action in the MSRDJ formalism comes at the cost of introducing an additional field $\wt{R}$ (referred to as \emph{response field}, as it encodes properties of response functions). In the MSRDJ action, the deterministic part of the Langevin equation \eqref{eq:Langevin} (i.e., everything but $\xi$) appears multiplied by the response field. Therefore, the effective action in our case reads
\be
	S = \int \rmd^d x  \rmd t \,  \wt{R} \lqq  (\partial_t - D\nabla^2 + u_2) R + u_3 R \mal{R}_1 - u_4 R \mal{R}_1^2 \textcolor{red}{+ c_3 R^2 - c_4 R \mal{R}_2} \rqq + (\text{higher orders in } \wt{R}),
	\label{eq:effact}
\ee
where the higher orders in $\wt{R}$ are generated by the properties of the noise $\xi$ and are therefore ``GEP-like''. One typically fixes the scaling dimension of the coordinates to be $\lqq \vec{x} \rqq = -1$, which implies, due to the diffusive nature of the problem, that $\lqq t \rqq = -2$ for time (in other words, the \emph{dynamic critical exponent} is $2$). Taking from \cite{Janssen1985} the scaling dimensions of the fields
\be
	\sdim{R} = \frac{d+2}{2} \mand \sdim{\wt{R}} =  \frac{d-2}{2},
\ee
we thus find
\be
	\sdim{\mal{R}_1} = \frac{d+2}{2} -2 = \frac{d-2}{2} \mand \sdim{\mal{R}_2} = 2\frac{d+2}{2} -2 = d .
\ee
we are now able to calculate the dimensions of all coupling simply by requiring each term of the action to have net dimension zero. As an example, we work out here $\sdim{u_3}$; we impose
\be
	0 = \sdim{\int \rmd^dx \rmd t \, u_3 \, \wt{R} R \mal{R}_1} = \underbrace{\sdim{\rmd^dx \rmd t}}_{-d-2} + \sdim{u_3} + \underbrace{\sdim{\wt{R}}}_{\frac{d-2}{2}} + \underbrace{\sdim{R}}_{\frac{d+2}{2}} + \underbrace{\sdim{\mal{R}_1}}_{\frac{d-2}{2}},
\ee
i.e., $\sdim{u_3} = 3 - d/2$, which is $>0$ for $d < 6$. Therefore, this term is only relevant in spatial dimensions smaller than $d_c = 6$, which represents the upper critical dimension for dynamic isotropic percolation. The other couplings' dimensions can be calculated in an analogous fashion, yielding
\be
	\sdim{u_4} = 4 - d \comma \sdim{c_3} = \frac{2-d}{2}  \mand \sdim{c_4} = 2-d.
\ee
These values are all $< 0$ at the upper critical dimension $d = d_c = 6$, showing that the corresponding terms in the action are strictly \emph{irrelevant} and only need to be accounted for if the coupling $u_3$ is fine-tuned to $0$. We have shown above, however, that this never happens in our equations, even if the more stringent constraint (B) is enforced.

We finally discuss the effect of an additional process which introduces recovery without immunization, i.e., a decay $\ket{r} \to \ket{g}$. Assuming it occurs uniformly throughout the system at a rate $\lambda$ the equations of motion get modified according to
\begin{eqnarray}
  \partial_t R&=& z\alpha R (G - cR) \mal{F} -\kappa R - \lambda R, \\
  \partial_t G&=& - z\alpha  R  (G - cR) \mal{F} + \lambda R,\\
  \partial_t N&=&\kappa R .
\end{eqnarray}
By defining $\mal{G} = G - \lambda / (z\alpha)$, however, we can recast them in the same form as before
\begin{eqnarray}
  \partial_t R&=& z\alpha R (\mal{G} - cR) \mal{F} -\kappa R , \\
  \partial_t G&=& - z\alpha  R  (\mal{G} - cR) \mal{F},\\
  \partial_t N&=&\kappa R .
\end{eqnarray}
Hence, one can follow exactly the same steps and find a Langevin equation like \eqref{eq:Langevin} up to the substitution $G_0 \to \mal{G}_0$. The introduction of this process thereby amounts to an overall rescaling of the couplings of the effective action \eqref{eq:effact}:
\be
	u_2 = \kappa - z\alpha \mal{G}_0, \quad u_3 = (z\alpha)^2 \mal{G}_0, \quad u_4 = \mal{G}_0 \frac{(z\alpha)^3}{2}.
\ee
The vanishing of the quadratic term occurs at $\mal{G}_0 = \kappa / (z\alpha)$, at which $u_3 = z\alpha \kappa >0$, implying that the new process does not affect the fundamental properties of the theory, i.e., the leading non-linearity at the critical point is still the cubic one. Of course, the actual position of the critical point is shifted to larger values of $\alpha$, as one could expect considering that this process hinders the growth of a self-sustaining front of infection (an outbreak).

\section{Repeating pattern of $N_{\rm{SS}}$ in the quantum case}

We give here the details on how the theoretical (blue) curve in Fig.~\ref{fig:3}(a) in the main text can be determined. We start from the inhomogeneous mean-field equations for the quantum case, which we reproduce below for the reader's convenience:
\begin{eqnarray}
  \partial_t R_m&=&\Omega \overline{R}_m \Sigma_m  -\kappa R_m, \label{eq:Q1-1}\\
  \partial_t G_m&=&-\Omega \overline{R}_m \Sigma_m ,\\
  \partial_t N_m&=&\kappa R_m,\\
  \partial_t \Sigma_m&=&-2\Omega \overline{R}_m(R_m - G_m) - \frac{\kappa}{2} \Sigma_m. \label{eq:Q4-1}
\end{eqnarray}
Firstly, by measuring energies in units of $\kappa$ and time in units of $\kappa^{-1}$ we can set, without loss of generality, $\kappa = 1$. Secondly, we neglect the third equation, since the remaining three close among themselves. Thirdly. we define the rescaled variables
\be
	R_m = \frac{Q_m}{\Omega}  \comma  G_m = \frac{F_m}{\Omega} \mand S_m = \frac{\Sigma_m}{\Omega}, 
	\label{eq:resc}
\ee
and consistently
\be
	\overline{R}_m = \frac{\overline{Q}_m}{\Omega} = \frac{1}{\Omega} \sum_{l \in \Lambda_m} Q_l.
\ee
With this simple rescaling, the equations read
\begin{eqnarray}
  \partial_t Q_m&=&  \overline{Q}_m S_m  - Q_m, \label{eq:R1}\\[1mm]
  \partial_t F_m&=& - \overline{Q}_m S_m ,\\
  \partial_t S_m&=&-2 \overline{Q}_m(Q_m - F_m) - \frac{1}{2} S_m \label{eq:R3} , 
\end{eqnarray}
showing no residual dependence on the driving parameter $\Omega$. A generic solution for $R_m$ of Eqs.~\eqref{eq:Q1-1}-\eqref{eq:Q4-1} will be a function of time, $\Omega$ and the initial conditions (denoted in the following with an additional subscript ``$0$''), i.e.
\be
	R_m \equiv R_m \lt t, \Omega, \set{R_{0k}}_k, \set{G_{0k}}_k, \set{\Sigma_{0k}}_k    \rt.
\ee
Correspondingly, for $Q_m$ and Eqs.~\eqref{eq:R1}-\eqref{eq:R3} we can write
\be
	Q_m \equiv Q_m \lt t, \set{Q_{0k}}_k, \set{F_{0k}}_k, \set{S_{0k}}_k    \rt.
\ee
Out of the rescaling transformation \eqref{eq:resc} we can thus connect the two expressions above according to
\be
	Q_m \lt t, \set{Q_{0k}}_k, \set{F_{0k}}_k, \set{S_{0k}}_k    \rt = \Omega  R_m \lt t, \Omega,   \set{\frac{Q_{0k}}{\Omega}}_k, \set{\frac{F_{0k}}{\Omega}}_k, \set{\frac{S_{0k}}{\Omega}}_k    \rt.
\ee
Via a completely analogous procedure, one also finds
\be
\begin{split}
	F_m \lt t, \set{Q_{0k}}_k, \set{F_{0k}}_k, \set{S_{0k}}_k    \rt = \Omega  G_m \lt t, \Omega,   \set{\frac{Q_{0k}}{\Omega}}_k, \set{\frac{F_{0k}}{\Omega}}_k, \set{\frac{S_{0k}}{\Omega}}_k    \rt, \\
	S_m \lt t, \set{Q_{0k}}_k, \set{F_{0k}}_k, \set{S_{0k}}_k    \rt = \Omega  \Sigma_m \lt t, \Omega,   \set{\frac{Q_{0k}}{\Omega}}_k, \set{\frac{F_{0k}}{\Omega}}_k, \set{\frac{S_{0k}}{\Omega}}_k    \rt.
\end{split}
\ee
Since the left-hand sides are independent of $\Omega$, the right-hand sides are as well. This implies that dynamic trajectories at different values of $\Omega$ can be identified up to an appropriate rescaling of the initial condition and of the overall amplitude, e.g.
\be
	\Omega  R_m \lt t, \Omega,   \set{\frac{Q_{0k}}{\Omega}}_k, \set{\frac{F_{0k}}{\Omega}}_k, \set{\frac{S_{0k}}{\Omega}}_k    \rt = \Omega'  R_m \lt t, \Omega',   \set{\frac{Q_{0k}}{\Omega'}}_k, \set{\frac{F_{0k}}{\Omega'}}_k, \set{\frac{S_{0k}}{\Omega'}}_k    \rt \quad \forall \,\, \Omega,\, \Omega',
\ee
which, re-expressing the initial condition in terms of the original variables, can be recast as
\be
	 \frac{\Omega}{\Omega'}  R_m \lt t, \Omega,   \set{\frac{\Omega'}{\Omega} R_{0k}}_k, \set{\frac{\Omega'}{\Omega}G_{0k}}_k, \set{\frac{\Omega'}{\Omega} \Sigma_{0k}}_k    \rt = R_m \lt t, \Omega', \set{R_{0k}}_k, \set{G_{0k}}_k, \set{\Sigma_{0k}}_k   \rt \quad \forall \,\, \Omega,\, \Omega'.
	 \label{eq:scaling}
\ee
Before proceeding, it is important to mention a consequence of the physical constraint $G_m + R_m + N_m = 1$ $ \forall m$: if a process at frequency $\Omega'$ starts from a vanishing density of immunes $N_{0k} \equiv 0$, the corresponding rescaled one at frequency $\Omega \neq \Omega'$ will instead feature a non-vanishing one
\be
	N_{0k} = 1 - \frac{\Omega'}{\Omega} (R_{0k} + G_{0k}) = 1 - \frac{\Omega'}{\Omega} \neq 0.
\ee

In order to apply the considerations above to the problem under study, we repeat here the main assumptions we make. First of all, from the numerical dynamics we have noticed that, to a reasonable approximation, every outbreak seems to produce a spatially uniform density of immune sites and, moreover, it appears that outbreaks always propagate from a very localized patch of infected sites in the very neighborhood of the center of the lattice. Hence, we impose that 
\begin{itemize}
	\item[(I)] The density of immunes after the $(j-1)$-th outbreak is $N_m \equiv N^{(j)}$ for all sites $m$ in an appropriate large neighborhood of the center.
	\item[(II)] Denoting for brevity with an index $\mathbf{k}$ the central site of the lattice, at some time $t_j$ after the $(j-1)$-th outbreak, the aforementioned large neighborhood in (I) reaches a configuration in which the center is the only site featuring a non-vanishing infected density $R_{\bf{k}} > 0$ and vanishing healthy density and coherence $G_{\bf{k}} = \Sigma_{\bf{k}} = 0$, whereas for all the others $R_m = \Sigma_m = 0$ and $G_m > 0$.
\end{itemize}
Furthermore, from the stationary data we have verified that the decreasing branch of $N_{\rm{SS}}$ after the first jump is reasonably well fitted by a power-law behavior $\propto 1 / \Omega$. Therefore, we further assume that
\begin{itemize}
	\item[(III)] The density $N^{(2)}$ left behind by the first outbreak obeys $N^{(2)} = A_1 / \Omega$ for some $A_1 > 0$ for all $\Omega$ such that a first outbreak is produced ($\Omega > \wt{\Omega}_1$ in the main text notation).
\end{itemize}

We now consider a process at frequency $\Omega$ which starts from the same initial conditions we employ throughout the main text, i.e.,
\be
\begin{split}
	&R_{0\bf{k}} = 1 \comma G_{0\bf{k}} = 0 \comma \Sigma_{0\bf{k}} = 0 \mand  \\
	&R_{0m} = 0 \comma G_{0m} = 1 \comma \Sigma_{0m} = 0,
\end{split}
\ee
and which produces at least $(j-1)$ outbreaks. In the following, for the sake of brevity we shall refer to these initial conditions as the ``standard'' ones. By (I) and (II), at some time $t_j$, in an approximately circular neighborhood of the center which reaches up to the $(j-1)$-th wavefront, the process satisfies
\be
\begin{split}
	&R_{\bf{k}}(t_j) = 1 - N^{(j)} \comma G_{\bf{k}} (t_j) = 0 \comma \Sigma_{\bf{k}}(t_j) = 0 \mand  \\
	&R_{m}(t_j) = 0 \comma G_{m} (t_j)= 1 - N^{(j)}  \comma \Sigma_{m}(t_j) = 0.  \label{eq:newin}
\end{split}
\ee
Since the equations of motion do not explicitly depend on time, we can reset our clock at $t_j$ and consider the subsequent dynamics as a new process which starts at $t = 0$ from initial conditions given by \eqref{eq:newin}, for which we thus introduce the notation $R_k (t_j) \to R_{0k}^{(j)}$, $G_k (t_j) \to G_{0k}^{(j)}$, $\Sigma_k (t_j) \to \Sigma_{0k}^{(j)}$ $\forall k$. We note now that
\be
	R_{0k}^{(j)} = \lt 1 - N^{(j)}  \rt R_{0k} \comma  G_{0k}^{(j)} = \lt 1 - N^{(j)}  \rt G_{0k} \mand  \Sigma_{0k}^{(j)} = \lt 1 - N^{(j)}  \rt \Sigma_{0k},
\ee
i.e.~the initial conditions of the process at $t_j$ are rescaled by $1 - N^{(j)} < 1$ with respect to the standard ones. Hence, we can now use Eq.~\eqref{eq:scaling} in order to identify a rescaled process which shows the same dynamical behavior up to multiplication by an overall amplitude and starts from standard initial conditions, evolving under a frequency which we rename $\Omega' \to \Omega^{(j)}$ (as in the main text) to keep in mind that the process we are interested in is evolving after the production of $(j-1)$ outbreaks. Comparing the arguments of the l.h.s.~of Eq.~\eqref{eq:scaling} with \eqref{eq:newin}, we find
\be
	\frac{\Omega^{(j)}}{\Omega} R_{0k} = R_{0k}^{(j)} = (1-N^{(j)}) R_{0k} \mand \frac{\Omega^{(j)}}{\Omega} G_{0k} = G_{0k}^{(j)} = (1-N^{(j)}) G_{0k},
\ee
implying
\be
	\Omega^{(j)} = \Omega (1- N^{(j)}) < \Omega,
	\label{eq:omegar}
\ee
as reported in the main text. Recalling the notation $\wt{\Omega}_j$ for the position of the $j$-th jump, we can draw now some predictions: if $\Omega^{(j)} < \wt{\Omega}_1$ the equivalent process evolving under $\Omega^{(j)}$ will not produce any outbreak. Consequently, the process we were originally interested in will neither, meaning that the process at fixed $\Omega$ will stop after $(j-1)$ outbreaks and produce a stationary density $N_{\rm{SS}} = N^{(j)}$ up to subextensive contributions generated by the finite propagation of the central infection, which becomes negligible in the thermodynamic limit. We thereby see that, by definition, $\Omega < \wt{\Omega}_j$. If instead $\wt{\Omega}_1 < \Omega^{(j)} < \wt{\Omega}_2$, the equivalent process at $\Omega^{(j)}$ will produce a single outbreak, after which it will stop. Hence, the original process will produce, after $t_j$, a $j$-th outbreak, but a $(j+1)$-th will not take place. This implies that $\wt{\Omega}_j < \Omega < \wt{\Omega}_{j+1}$. This identification can be further extended to any number of outbreaks. By recalling that $N^{(j)}$ is the density of immunes left behind by the $(j-1)$-th outbreak taking place in a process of frequency $\Omega$, and is thus a function of $\Omega$ itself ($N^{(j)} = N^{(j)} (\Omega)$), we can also write the relation
\be
	\wt{\Omega}_{i} = \wt{\Omega}_{i+j-1} \lt 1- N^{(j)} \lt \wt{\Omega}_{i+j-1}\rt \rt.
	\label{eq:rel0}
\ee
between the positions of the jumps. Since the l.h.s.~does not depend on $j$, this is easily generalized to
\be
	\wt{\Omega}_{i+j-1} \lt 1- N^{(j)} \lt \wt{\Omega}_{i+j-1}\rt \rt = \wt{\Omega}_{i+j'-1} \lt 1- N^{(j')} \lt \wt{\Omega}_{i+j'-1}\rt \rt  \quad \forall \, i,\,j,\, j'.
	\label{eq:w1}
\ee
We specialise this to $i = 0$
\be
	\wt{\Omega}_{j-1} \lt 1- N^{(j)} \lt \wt{\Omega}_{j-1}\rt \rt = \wt{\Omega}_{j'-1} \lt 1- N^{(j')} \lt \wt{\Omega}_{j'-1}\rt \rt  \quad \forall \,j,\, j',
	\label{eq:w2}
\ee
and to $i = 1$
\be
	\wt{\Omega}_{j} \lt 1- N^{(j)} \lt \wt{\Omega}_{j}\rt \rt = \wt{\Omega}_{j'} \lt 1- N^{(j')} \lt \wt{\Omega}_{j'}\rt \rt  \quad \forall \,j,\, j'.
\ee
Now, we recall that --- up to subextensive contributions which vanish in the thermodynamic limit --- $N^{(j)} (\Omega) = N_{\rm{SS}} (\Omega)$ $\forall \,\, \wt{\Omega}_{j-1} < \Omega < \wt{\Omega}_j$. Calling $\wt{N}_j^{(\pm)} = N_\mathrm{SS} (\Omega \to \wt{\Omega}_j^\pm)$ the top ($+$) and bottom ($-$) of the $j$-th jump, we find from Eqs.~\eqref{eq:w1} and \eqref{eq:w2} that
\be
	\wt{\Omega}_j \lt 1 -\wt{N}_j^{(\pm)}   \rt = \wt{\Omega}_{j'} \lt 1 -\wt{N}_{j'}^{(\pm)}   \rt \quad \forall \,j,\, j'.
	\label{eq:jump_pos}
\ee
These relations, extended to generic $\Omega$s, identify the enveloping black curves in Fig.~\ref{fig:3}(a) in the main text. More precisely, using the first jump as a reference, and recalling that $N_1^{(-)} = 0$, we find the expressions used in the plot
\be
	\Omega (1 - N) = \wt{\Omega}_1  \mand \Omega (1 - N) = \wt{\Omega}_1 ( 1 - N_1^{(+)} )
	\label{eq:envelopes}
\ee
where we numerically estimated $\wt{\Omega}_1 \approx 0.8571 \kappa$ (where we have reinstated the units $\kappa$) and $N_1^{(+)} \approx 0.898$.

We now make use of assumption (III) as well. First of all, this tells us that $N_{\rm{SS}} (\Omega) = A_1 / \Omega$ $\forall \,\, \wt{\Omega}_1 <\Omega < \wt{\Omega}_2$, which allows us to fix the constant $A_1$ in such a way that the fitting curve passes through the tip of the first jump: $A_1 = \wt{\Omega}_1 \wt{N}_1^{(+)}$. We now proceed by induction: we assume that $N^{(j)} (\Omega) = A_{j-1} / \Omega$ $\forall \,\, \Omega > \wt{\Omega}_{j-1}$ --- which corresponds to (III) for $j = 2$ --- and we wish to show the same is true for $N^{(j+1)}$. To this end, we consider a process at frequency $\Omega$ after its $(j-1)$-th outbreak started, such that the density of immunes in the proximity of the center is $N^{(j)} = A_{j-1}/ \Omega$. According to Eq.~\eqref{eq:omegar}, its equivalent process will take place at a rescaled frequency 
\be 
	\Omega^{(j)} = \Omega (1 - N^{(j)}) = \Omega - A_{j-1}.
	\label{eq:step0}
\ee
%
We now rewrite Eq.~\eqref{eq:scaling} as
\be
\begin{split}
	\Omega R_m \lt t, \Omega, \set{R_{0k}^{(j)}}_k, \set{G_{0k}^{(j)}}_k, \set{\Sigma_{0k}^{(j)}}_k  \rt = \Omega^{(j)}  R_m \lt t, \Omega^{(j)}, \set{R_{0k}}_k, \set{G_{0k}}_k, \set{\Sigma_{0k}}_k \rt. \\
	\Omega G_m \lt t, \Omega, \set{R_{0k}^{(j)}}_k, \set{G_{0k}^{(j)}}_k, \set{\Sigma_{0k}^{(j)}}_k  \rt = \Omega^{(j)} G_m \lt t, \Omega^{(j)}, \set{R_{0k}}_k, \set{G_{0k}}_k, \set{\Sigma_{0k}}_k \rt.
	\label{eq:GG}
\end{split}
\ee
Since $R_m + G_m + N_m = 1$ at all times, we can also write
\be
	\Omega \lqq  1 - N_m \lt t, \Omega, \set{R_{0k}^{(j)}}_k, \set{G_{0k}^{(j)}}_k, \set{\Sigma_{0k}^{(j)}}_k  \rt \rqq = \Omega^{(j)} \lqq 1 -  N_m \lt t, \Omega^{(j)}, \set{R_{0k}}_k, \set{G_{0k}}_k, \set{\Sigma_{0k}}_k \rt \rqq.
\ee
Now, the first outbreak produced by the process in the r.h.s.~corresponds to the $j$-th one for the process in the l.h.s., leading to the identity
\be
	\Omega \lt 1 - N^{(j+1)} \lt \Omega \rt  \rt  = \Omega^{(j)} \lt  1 - N^{(2)}  \lt \Omega^{(j)} \rt    \rt
\ee
for the densities of immunes left behind by either. Since by (III) we know that
\be
	N^{(2)}  \lt \Omega^{(j)} \rt = \frac{A_1}{\Omega^{(j)}},
\ee
this implies 
\be
	\Omega \lt 1 - N^{(j+1)} \lt \Omega \rt  \rt = \Omega^{(j)} - A_1.
\ee
By the inductive step in the formulation \eqref{eq:step0}, we then arrive at
\be
	\Omega \lt 1 - N^{(j+1)} \lt \Omega \rt  \rt = \Omega - A_{j-1} - A_1,
\ee
or equivalently
\be
	N^{(j+1)} \lt \Omega \rt  = \frac{A_1 + A_{j-1}}{\Omega},
\ee
which, with the identification $A_j = A_1 + A_{j-1}$, concludes our inductive proof. In particular, this also tells us that $A_j = j A_1 = j \wt{\Omega}_1 \wt{N}^{(+)}_1$.

Now, exploiting the fact that $N^{(j+1)} \lt \Omega \rt = N_{\rm{SS}} \lt \Omega \rt$ $\forall \,\,\wt{\Omega}_{j} < \Omega < \wt{\Omega}_{j+1}$, we find
\be
	N_{\rm{SS}} \lt \Omega \rt = \sum_{j=1}^\infty \theta \lt \Omega - \wt{\Omega}_{j} \rt \, \theta\lt   \wt{\Omega}_{j+1} - \Omega \rt  \frac{j A_1}{ \Omega} = \sum_{j=1}^\infty \theta \lt \Omega - \wt{\Omega}_{j} \rt \, \theta\lt   \wt{\Omega}_{j+1} - \Omega \rt  \frac{j \wt{\Omega}_1 \wt{N}^{(+)}_1}{ \Omega},
\ee
where $\theta$ denotes the Heaviside step function ($\theta(\omega > 0 ) = 1$, $\theta(\omega < 0) = 0$). The expression above corresponds to the blue line in Fig.~\ref{fig:3}(a) in the main text.

The considerations above also allow us to estimate the threshold frequencies $\wt{\Omega}_j$ for $j > 1$, given the properties of the first jump, i.e.~the values $\wt{\Omega}_1$ and $\wt{N}_1^{(+)}$. These values correspond to the points where the (blue) theoretical curve intersects the (black) enveloping ones \eqref{eq:envelopes}. Setting $i=1$ in Eq.~\eqref{eq:rel0}, we find
\be
	\wt{\Omega}_1 = \wt{\Omega}_j \lt  1 - N^{(j)} \lt \wt{\Omega}_j  \rt  \rt = \wt{\Omega}_j \lt 1 - \frac{(j-1) \wt{\Omega}_1 \wt{N}_1^{(+)}}{\wt{\Omega}_j}   \rt,
\ee
i.e.
\be
	\wt{\Omega}_j = \wt{\Omega}_1 \lqq  1 + (j-1) \wt{N}_1^{(+)}  \rqq.
\ee



\end{document}